\documentclass[traditabstract]{aa}
\usepackage{aas_macros}
\usepackage{hyperref}
\hypersetup{colorlinks=true,citecolor=blue}
\pdfoutput=1

\usepackage[latin1]{inputenc}
\usepackage{amsmath}
\usepackage{amsfonts}
\usepackage{amssymb}

\usepackage{url} 
\usepackage{longtable}
\usepackage{graphicx}
\usepackage{txfonts}
\usepackage{natbib}
\usepackage{xspace}
\usepackage{color}
\usepackage{hyperref}
\usepackage{multirow}
\usepackage{todonotes}

\usepackage{appendix}
\usepackage{natbib} 

\bibpunct{(}{)}{;}{a}{}{,}

\begin{document}

\title{The PCA Lens-Finder: application to CFHTLS}
\author{D. Paraficz\inst{\ref{EPFL}}, F. Courbin\inst{\ref{EPFL}}, A. Tramacere\inst{\ref{unige}}, R. Joseph\inst{\ref{EPFL}}, R.B. Metcalf\inst{\ref{Bologna}}, J.-P. Kneib\inst{\ref{EPFL}}, P. Dubath\inst{\ref{unige}}, D. Droz\inst{\ref{EPFL}}, F. Filleul\inst{\ref{EPFL}}, D. Ringeisen\inst{\ref{EPFL}}, C. Sch{\"a}fer\inst{\ref{EPFL}}
 }

\institute{Laboratoire d'Astrophysique
Ecole Polytechnique F\'ed\'erale de Lausanne (EPFL)
Observatoire de Sauverny
CH-1290 Versoix \label{EPFL}
\and
Department of Astronomy, University of Geneva, Ch. d'\'Ecogia 16, CH-1290 Versoix, Switzerland \label{unige}
\and
Dipartimento di Fisica e Astronomia - Universita di Bologna, via Berti Pichat 6/2, 40127, Bologna, Italy \label{Bologna}
}
\date{\today}
\authorrunning{Paraficz}
\abstract{We present the results of a new search for galaxy-scale strong lensing systems in CFHTLS {\it Wide}.
Our lens-finding technique involves a preselection of potential lens galaxies, applying simple cuts in size and magnitude. We then perform a Principal Component Analysis of the galaxy images, ensuring a clean removal of the light profile. Lensed features are searched for in the residual images using the clustering topometric algorithm \textsc{DBSCAN}. We find 1098 lens candidates that we inspect visually, leading to a cleaned sample of 109 new lens candidates. 
Using realistic image simulations we estimate the completeness of our sample and show that it is independent of source surface brightness, Einstein ring size (image separation) or lens redshift. We compare the properties of our sample to previous lens searches in CFHTLS. 
 Including the present search, the total number of lenses found in CFHTLS amounts to 678, which corresponds to $\sim$4 lenses per square degree down to $i(AB)=24.8$. This is equivalent to $\sim$ 60.000 lenses in total in a survey as wide as Euclid, but at the CFHTLS resolution and depth.}
 
\maketitle

\section{Introduction}

%=================================================================================================================

Strong gravitational lensing occurs when light rays emitted by a distant source are deflected by the potential well of a foreground mass \citep{Einstein:1916}. If the latter is compact enough  i.e. above the lensing critical surface mass density, multiple images of the background source are formed. Because strong lensing has only simple dependence on its geometry and fundamental physics (general relativity),  thus its applications in cosmology and in the study of galaxy formation and evolution are straightforward and becoming more and more numerous. 

Strong gravitational systems have been used in  recent years to address key scientific questions. 
In particular, strong lensing consists of a powerful tool to map the total mass (dark and luminous) in galaxies, independently of their dynamical state or star formation history \citep[e.g.][]{Treu:2002a, Rusin:2003, Treu:2004, Rusin:2005, Sonnenfeld:2012, Bolton:2012}. Also, thanks to strong lensing, small and dark satellite galaxies have even been found and weighted \citep[e.g.][]{Metcalf:2001, Dalal:2002,  Treu:2004, Koopmans:2006, Jiang:2007, Gavazzi:2007, Treu:2010, Auger:2010, Bolton:2012, Sonnenfeld:2012, Vegetti:2012,  Oguri:2014}. Applications in cosmology using the time delay method \citep[e.g.][]{Refsdal:1964} between the  multiply-lensed images of distant quasars are also becoming of increasing interest thanks to intensive photometric monitoring programs like COSMOGRAIL  \citep[e.g.][]{Vuissoz:2008, Courbin:2011, Rathna:2013, Tewes:2013}.  In combination with state-of-the-art modelling tools, these time delays can be used to constrain the cosmological parameters  both with precision and accuracy \citep[e.g.][]{Suyu:2012, Suyu:2010, Suyu:2013, Suyu:2014}. Even without time delay measurements, large samples of galaxy-scale strong lenses can help to constrain cosmology, as  \cite{cao2015} did, using 118 systems to constrain the dark energy equation of state parameter, $w$. 

The above applications of strong lensing are possible because: 1. significantly large samples of strong lensing systems are now available, 2. some of the discovered systems have specific, rare properties making them particularly effective in delivering astrophysical or cosmological constraints. The ongoing  (DES, KIDS) and planned wide field surveys of the next decade  (Euclid, LSST, WFIRST) will continue to revolutionise the field, by making available hundreds of thousands of new strong lenses \cite[e.g.][]{Pawase:2014, Collett:2015}, i.e. $\sim$3 orders of magnitude larger than the current sample sizes. 

Early systematic searches for strong lenses took advantage of the so-called lensing magnification bias, i.e. the fact that a lensed source is seen brighter because it is lensed.  These source-selected lensing system samples were built by looking for multiple images among samples of optically bright quasars \citep[e.g.][]{Surdej:1987, Magain:1988}. This was followed up in a more systematic way with a search in the Hamburg-ESO bright quasar catalogue \citep{Wisotzki:1993, Wisotzki:1996, Wisotzki:1999, Wisotzki:2002, Wisotzki:2004, Blackburne:2008}, in the SDSS with the Sloan Quasar Lens Survey  \citep[SQLS;][]{Inada:2003, Inada:2012, Oguri:2006, Oguri:2008,Inada:2007} as well as in other wide-field optical observations \citep[e.g.][]{Winn:2000, Winn:2001, Winn:2002a, Winn:2002b}. Similarly, strong lens searches were also carried out in the radio in the FIRST survey \citep[][]{Gregg:2000} and in the CLASS survey \citep[][]{Myers:2003, Browne:2003}. More recently, the same strategy was adopted at millimeter wavelengths with the South Pole Telescope \citep[SPT][]{Hezaveh:2013}, and at sub-millimeter wavelengths with the Herschel satellite like H-ATLAS \citep{Negrello:2010, Gonzalez-Nuevo:2012, Bussmann:2013} and HerMES \citep{Conley:2011, Gavazzi:2011, Wardlow:2013}.

Source-selected samples of lensing systems are mostly composed of lensed quasars. Searches for non-quasar lensed galaxies are generally carried out by preselecting a sample of potential lensing galaxies and by looking for lensed images or spectra  in their background  \citep{Ratnatunga:1999, Fassnacht:2004, Moustakas:2007, Cabanac:2007, Belokurov:2007, Faure:2008, Marshall:2009, Pawase:2014, More:2015}. 
These lens-selected samples are best constructed using spectra where sets of emission lines at two (or more) distinct redshifts are looked for. The method was pioneered by \citep[][]{Willis:2005, Willis:2006} with their Optimal Line-of-Sight Survey, soon followed by the SloanLens ACS Survey \citep[SLACS, e.g.,][]{Bolton:2006, Treu:2006, Koopmans:2006, Gavazzi:2007, Gavazzi:2008, Bolton:2008, Auger:2009} and by the BOSS Emission-Line Lens Survey \citep[BELLS;][]{Brownstein:2012}. The \textsc{SLACS} sample on its own has about 100 confirmed gravitational lenses in the redshift range $0.1< z<0.4$ with HST imaging \citep[e.g.][]{Bolton:2006, Auger:2009}. 
The main advantage of the spectroscopic approach is that the redshifts of the lens and of the source are readily obtained, along with the stellar velocity dispersion in the lens \citep[e.g.][]{Treu:2004, Koopmans:2006, Auger:2010}. Moreover, if the source has strong emission lines, then the light  from the lens and the source can  easily be separated.

In the imaging, on the other hand, the source is often hidden in the lens glare, thus it cannot be detected so easily.  For this reason carrying out  an imaging lens search  requires careful image processing to efficiently remove the lens light and unveil any faint background lensed galaxy. Such techniques start to be implemented, and will become increasingly important with the development of large sky surveys like DES, KIDS, Euclid, the LSST and WFIRST. 

Based on two-band imaging, \citet{Gavazzi:2014} have devised a method to detect faint blue arcs behind foreground redder galaxies. 
They extend their technique to multi-band lens modelling \cite[][]{Brault:2015} 
and they apply them to the CFHTLS (Canada-France-Hawaii Telescope Legacy Survey) data \citep[][]{Cuillandre:2012}. 
A second method was introduced by \mbox{\citet{Joseph:2014}} that can work both in single-band and multi-band. 
It is based on a principal component analysis \citep[PCA;][]{Jolliffe:1986} of the full imaging dataset to subtract the image of galaxies, even in the presence of complex structures. The residual image can then be used to search for lensed sources. In this paper, we use the method of \citet{Joseph:2014} to extend the sample of  known galaxy-scale strong lenses in CFHTLS.

The paper is organized as follows. In Sect. 2, we provide a brief description of the observational material  and of the sample selection technique from the object catalogues for the CFHTLS. 
In Sect. 3, we describe the lens-finding algorithm based on PCA and its improvements. In Sect. 4, we present the list of our new lens candidates and compare it to previous results from other lens searches in the same area of sky, i.e., the CFHTLS fields. In Sect. 5,  we discuss the completeness of the sample based on lens simulation. Finally, in Sect. 6, we provide a summary of the main conclusions from this work and provide a short outlook for future progress.

Throughout this work, we assume $\Omega_0 = 0.3 $,  $\Omega_\Lambda = 0.7 $,
and  $H_0 = 70 $  ${\rm km s}^{-1} {\rm Mpc}^{-1}$. All magnitudes are in the AB
system \citep{Oke:1983}.

%%%%%%%%%=================================================================

%===============================
\section{Observational material and sample selection}
%===============================

Our main goal in this work is to complement and extend current samples of galaxy-scale strong lens candidates, starting  with the public imaging data from the CFHTLS. To do so, we use the new technique proposed by \citet{Joseph:2014}. 

%===============================
\subsection{CFHTLS data}
%===============================

\begin{figure}[t!]
\begin{center}
\includegraphics[scale=0.4,angle=0]{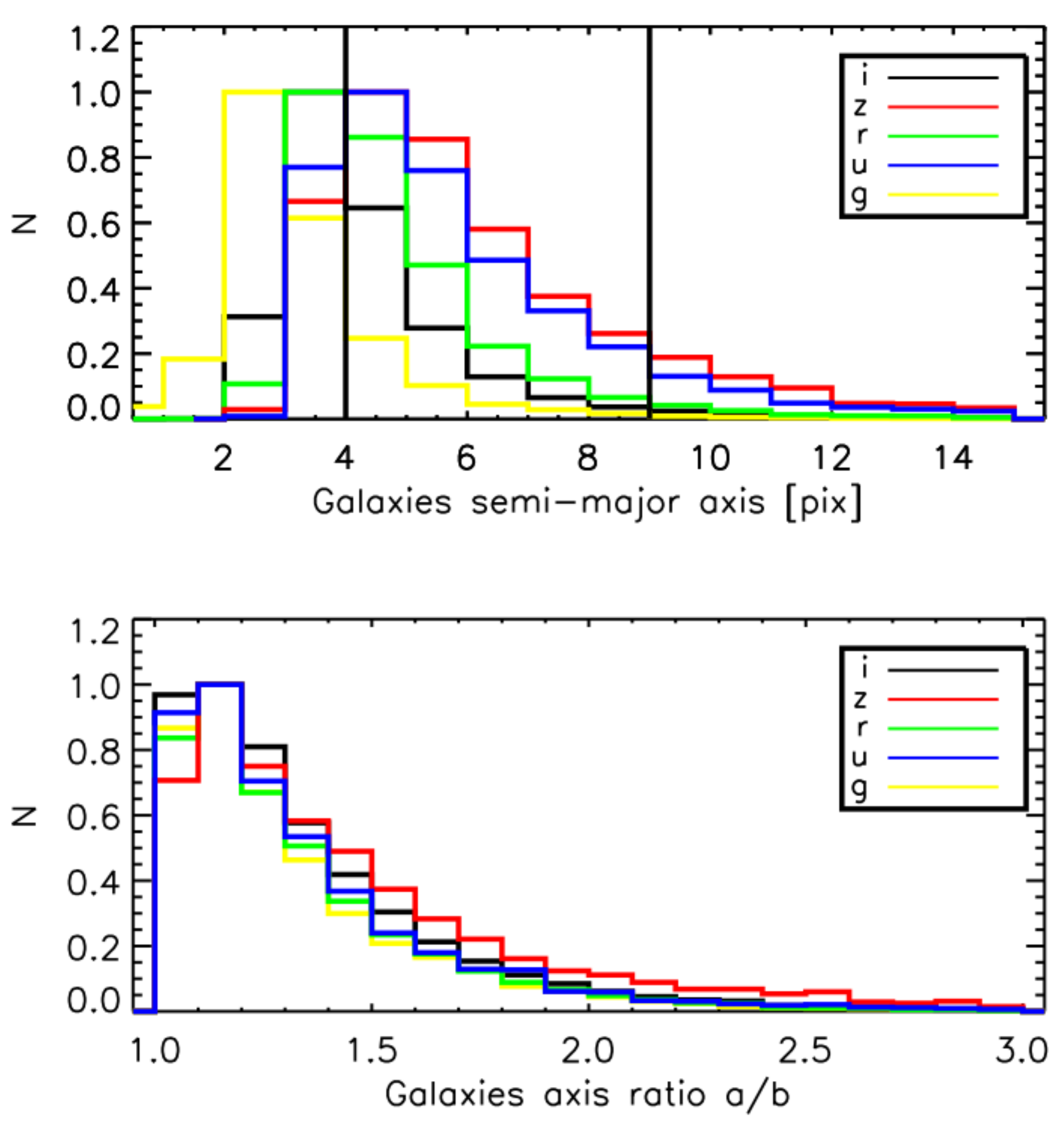}
\caption{Distribution of the semi-major axes and of their ratio, for our  preselections of galaxies  in the full CFHTLS sample \citep{Cuillandre:2012}. The vertical lines mark our size cut-off of galaxies. The pixel size is 0\farcs187.}
\label{a-distrib}
\end{center}
\end{figure}

\begin{figure*}
\hspace{-1.0cm}
\includegraphics[width=1.05\textwidth]{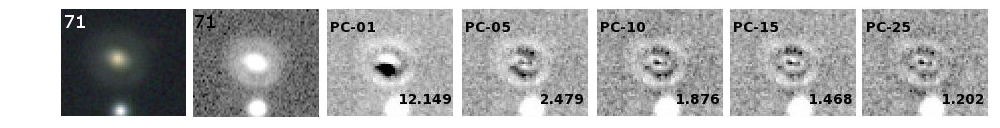}  
\caption{Illustration of the PCA reconstruction of a galaxy image. The first two panels show the original image (colour and single-band). Each of the other panels shows the residual image between the data and the reconstructed galaxy using respectively 1, 5, 10, 15 and 25 principal components (PCs). The value of the corresponding reduced $\chi^2$ is given in the lower right corner.}
\label{PCA}
\end{figure*}

The Canada-France-Hawaii-Telescope Legacy Survey is a large program consisting of 500 observing nights between January 2003 and early 2009, divided into two parts. The {\it Deep} survey has 4 ultra-deep pointings and the {\it Wide} survey has 171 intermediate-depth pointings. Because strong lensing systems are rare, we need to use the widest possible survey area, i.e. the {\it Wide} part of CFHTLS. 

The  {\it Wide} CFHTLS  \citep[][]{Cuillandre:2012}  covers 155 deg$^2$ divided  into four independent fields observed in  five bands. The limiting point-source AB magnitude (80\% completeness) for each band is $u = 25.2$, $g = 25.5$, $r = 25.0$, $i = 24.8$, $z = 23.9$. For each band the mean seeing is FWHM$=0\farcs85$, $0\farcs79$, $0\farcs71$, $0\farcs64$, and $0\farcs68$, respectively. The pixel size of the CFHTLS data  is 0\farcs187.

 In this paper, we use the T0007\footnote{See details at \url{http://www.cfht.hawaii.edu/Science/CFHLS/}}   final release of the CFHTLS with improved flat-fielding and photometric calibration techniques developed by the Supernova Legacy Survey (SNLS) team in collaboration with the CFHT. This release benefits from two types of  photometric catalogues: 1. source catalogues for individual images (i.e. the {\tt .ldac} files in the T0007 package) that we use to build our samples of lens galaxies in each filter separately, 2. merged source catalogues produced  from the $g$, $r$, and $i$ images that we use to infer the colour information of our lenses. The full description of the CFHTLS-T0007 release can be found in \citet[][]{Cuillandre:2012}.

%===============================
\subsection{Sample selection}
\label{selection}
%===============================

We preselect galaxies among the full CFHTLS source catalogues for individual images produced by \textsc{Terapix} using \textsc{SExtractor} \citep{Bertin:1996}. PCA  requires uniform in size, morphology and brightness sample of elliptical galaxies, and since   size and morphology of a galaxy changes between different bands, thus we create five catalogues, one for each of the five CFHTLS filters,  independently. A given galaxy can therefore appear in several of the catalogues. Before using the PCA subtraction technique of \citet{Joseph:2014}, we apply the following selection criteria to each of the five catalogues:

\begin{itemize}
\item We use only the objects classified as non-stellar by \textsc{SExtractor} (\textsc{CLASS}\textunderscore\textsc{STAR} $>$ 0.98) and we apply a detection threshold of 10$\sigma$ for each band, that is $u = 23.9$, $g = 24.3$, $r = 23.5$, $i = 23.5$, $z = 22.4$. Fainter objects would make difficult targets for  future spectroscopic follow-up. 
\item We apply a cut on the semi-axis ratio, $a/b < 3$, which includes most of the early-type galaxies (Park et al. 2007), but  rejects most  of the spurious objects like those ``created'' by diffraction patterns of bright stars. 
\item We apply an (angular) size cut-off. Small galaxies, with a semi-major axis $a<4$ pixels are excluded. Any arc in their vicinity can be detected without subtracting the light of the foreground galaxy. We also remove galaxies with $a>9$ pixels. These galaxies are rare  and therefore poorly modeled with the PCA technique \cite[][]{Joseph:2014}.  Since we want to ensure uniformity of galaxy shapes in a group, galaxy sizes are computed separately  in each band. Our final selection therefore spans sizes in the range $4<a<9$ pixels. Figure~\ref{a-distrib} displays the distribution in semi-major axis for the full sample in all the CFHTLS bands.
\end{itemize}

This leaves us with a pre-selection of early-type and late-type galaxies. However, spiral arms can be mistakenly taken for lensed arcs, resulting in false positives. To avoid this, we further restrict the sample to only  elliptical galaxies. This can be achieved by either using a galaxy classifier based in morphological features in images (e.g. the \texttt{ASTErIsM} software by A. Tramancere, et al., 2015 submitted) or by applying a colour selection. We adopt the latter strategy, selecting galaxies with $(g - i) > 1.0$ within a $3\arcsec$ aperture, following \citet{Gavazzi:2010}. Obviously, some of the potential lenses are missed by this selection, but this is the price to pay to remove  spiral galaxies efficiently.

For each selected object, we create an image stamp centred on the galaxy. Since rotation is not a principal component, we also apply a rotation to each stamp  to align the major axes of all galaxies. In doing so, we use a polynomial transformation and a bilinear interpolation. We note that we do not apply any other re-scaling. Instead, to ensure final uniformity of the PCA basis,  we take advantage of the very large sample  and  we split it in five bins of galaxies sizes.   The five groups are defined by  the  galaxies semi-major axis as follows:  (1): $a=[4-5]$, (2):~$a=[5-6]$, (3):~$a=[6-7]$, (4):~$a=[7-8]$, and (5):~$a=[8-9]$ pixels.

%===============================
\section{Lens-finding algorithm and improvements}
%===============================

Our algorithm by nature finds bright lensing galaxies. In such samples the lensed source is often hidden in the glare of the foreground galaxy, which must be properly removed before any search for lensed structures can be carried out. Our PCA-based lens-finder therefore includes two steps: 1. subtraction of the central galaxy from the original images/stamps, using PCA image reconstruction and 2. detection of lensed extended objects (arcs, rings) in the residual images.

\begin{figure}
\begin{center}
\includegraphics[scale=0.22,angle=0]{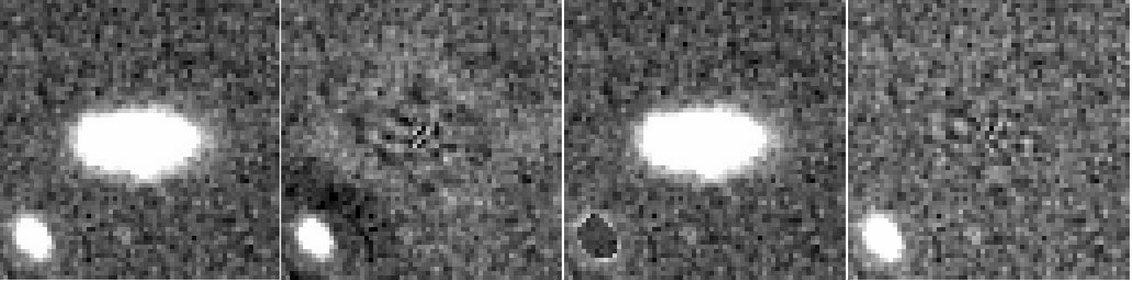}
\caption{Illustration of our masking strategy. {\it From left to right}: a) image of a galaxy from our sample, b) subtraction without using a mask during the reconstruction process, leaving ring-like artefacts, c) masked image used for the reconstruction process, d) resulting residual image, without any artefact.}
\label{mask}
\end{center}
\end{figure}

%===============================
\subsection{Removal of the lensing galaxy: the PCA approach}
%===============================

Traditional ways of subtracting galaxies  in imaging data are either to fit a    elliptical profile to the data with, for example, the {\tt galfit} software \citep{Peng:2010, Cabanac:2008} or to subtract aperture-scaled images in two different bands \citep{Gavazzi:2014}. As galaxies are not perfect elliptical profiles, these approaches often leads to significant flux residuals that prevents the detection of faint background lensed objects and they produce large amounts of false positives. 

Our solution to this problem is presented in \citet{Joseph:2014} \footnote{PCA script available at \url{https://github.com/herjy/PiCARD}}, where we build an empirical galaxy light model from the sample of galaxies itself using a principal component analysis. PCA decomposition of a dataset allows one to recognise any similarity among its elements: the elements in the dataset are converted into another set of variables called principal components (PC), which are orthogonal and ordered so that the first PC has the largest possible variance, the second PC has the second largest variance, and so on. The details of our PCA technique are provided in \citet{Joseph:2014}.

\begin{figure*}[t!]
\hspace{-1cm}
			\includegraphics[width=1.1\textwidth]{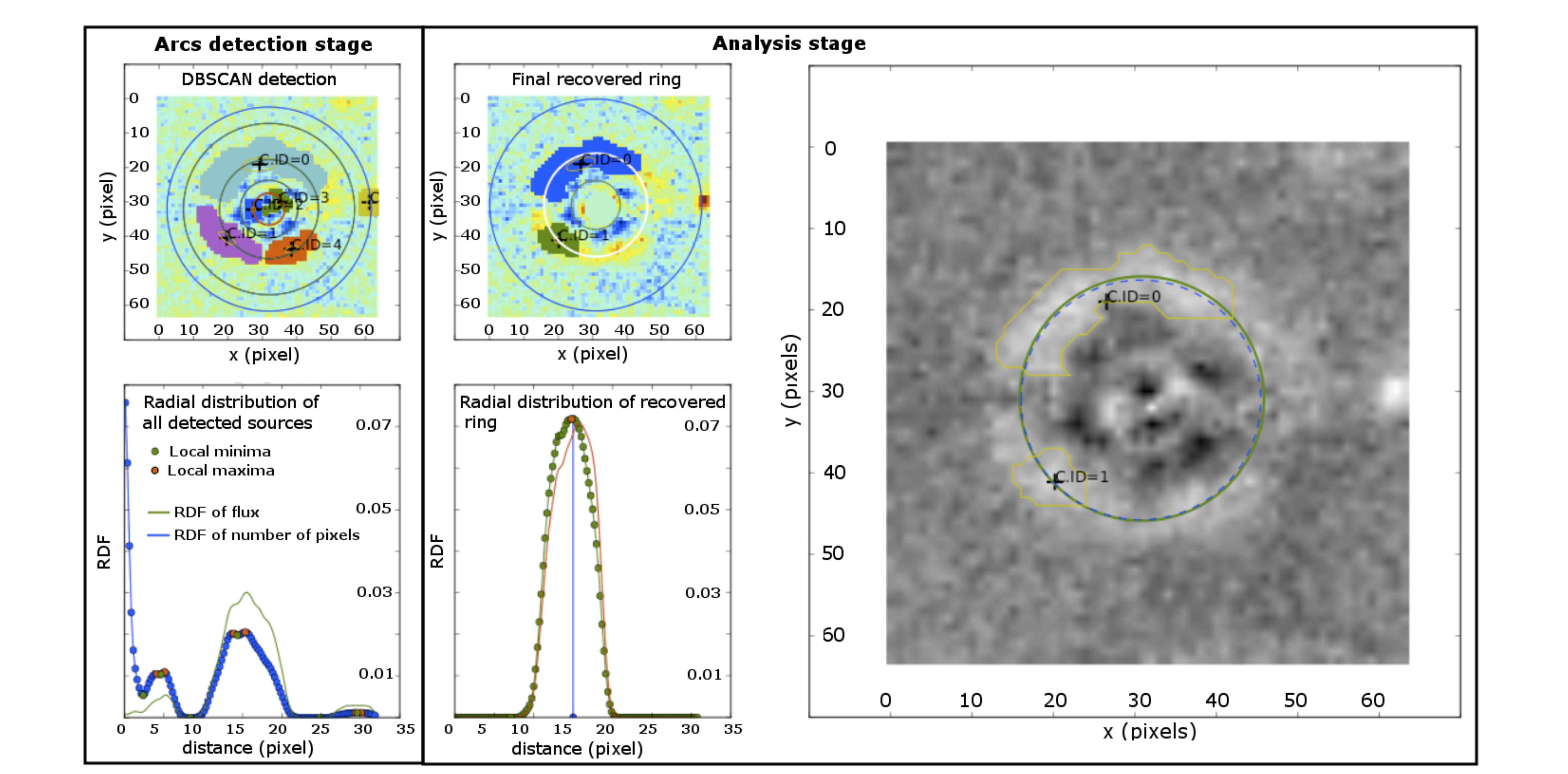}
	\caption {Analysis of the PCA-subtracted images. {\bf Arc detection stage.} \textit{ Top panel}: \texttt{DBSCAN} detection 
		of the sources.  \textit{Bottom panel}:    radial distribution of the detected  
		sources. The green solid line represents the pixel flux radial distribution, while the 
		blue line represents the radial distribution of the number of pixels in the sources.		
		{\bf Analysis stage.} \textit{Top panel}:  All sources smaller than a critical angular size (e.g. the PSF size) are removed, the  remaining sources are  merged together in the final ring  (indicated with white line). \textit{Bottom panel}
		The red solid line represents the pixel flux radial distribution, while the green line represents the radial distribution of the number of pixels in the sources.
		\textit{Right panel}: Residual image, where the green solid line shows the best circle fit to the final ring. The yellow lines show the contour of the components of the final ring.}
		\label{fig:DBSCAN}
\end{figure*}

A critical step in the PCA reconstruction is the choice of the number of PC coefficients. If all coefficients are used, the reconstructed image is  identical to the original image. This clearly leads to overfitting of the data and noise and will simply remove all structures of interest, like faint lensed rings and arcs. To circumvent this problem,  the galaxy image  needs to reconstructed using only a limited number of coefficients.  Obviously, there might be an optimal number of coefficients to be used to avoid over-fitting or under-fitting of the data. This optimal number of PC depends on the diversity in shape among the galaxies in the sample, i.e. the range in galaxy sizes, the presence of companions near the galaxies used to build the PCA basis, and it also depends on the number of objects used to build the PCA basis.  Fig.~\ref{PCA} gives an illustration of a galaxy image  reconstructed from a different  number of PCs. To evaluate the overall quality of reconstruction of a galaxy image in an objective way, we compute the reduced $\chi^2$ in the image after subtraction of the galaxy \citep[Eq. 6 in][]{Joseph:2014}. An ideal reconstruction gives a reduced $\chi^2$ close to one. Of course lensed features in the image do produce signal in the residual image,  a cut-off in $\chi^2$  has to be defined that ensures an adequate tradeoff between clean removal of the foreground galaxy and non-removal of any potential lensed feature.  Finding this cut-off is a subjective part of our procedure and final result for false to true positives ratio will strongly depend on the adopted value. In the present case we choose $\chi^2\sim1.4$ for the CFHTLS data (see Fig.~\ref{PCA}) and for the simulations used to evaluate completeness of the procedure  (see Section 5).

The PCA technique described in \citet{Joseph:2014} works well for isolated galaxies. In practice, however, galaxies often have companions, either physical or the result of  line-of-sight effects. Companions affect the results both when building the PCA basis from the galaxy sample and when reconstructing the image of a given galaxy. The first problem \citep[as described already in][]{Joseph:2014} is easily avoided by computing the PCA basis on a subset of galaxies with no bright companions. The large size of our galaxy sample allows us to do that in practice. However, companions are present in the images of the galaxy we want to reconstruct and subtract.  In the case of reconstruction of an image of a galaxy with bright companions, residuals might feature artifacts that mimic a ring (Fig.~\ref{mask}). To avoid this problem we  simply mask bright companion before the reconstruction process. Images reconstructed in this manner are then subtracted from original, non-masked images creating residuals that are  now free of the false rings. Figure~\ref{mask} illustrates the improvement over a non-masked image.

To apply the masking to all the companion sources in the image stamps, we use the \texttt{DBSCAN} algorithm implemented in the \texttt{ASTErIsM} software (Tramacere et al 2015 submitted). We identify the object at the centre of the stamp, as our source of interest that will be not masked. We iterate among all the remaining sources, and we mask all the sources with an integrated flux larger than the half of the central galaxy's integrated flux and all the sources with a distance from the central galaxy smaller than the half of the central galaxy's radius. The masked pixel fluxes are replaced with flux values randomly sampled from the background pixels flux distribution.

\subsection{Looking for lensed features in the residual image}

Once the galaxies have been properly removed from all preselected image stamps, we can now 
search for lensing features in these residual images.  To avoid too many false positives we choose to search for only  arc-like features. This was done using the cluster/island detection algorithm described in detail and tested in Tramacere et al (2015, in prep). Below, we provide  a short overview of the procedure. 
This method is based on the application of the  {\tt DBSCAN} clustering topometric algorithm  \footnote{  {\tt DBSCAN}  algorithm  available at a \citep{Ester96DBScan,Tramacere:2013it}} \citep{Ester96DBScan,Tramacere:2013it}, which extracts \textit{sources} in image stamps by defining density-based \textit{clusters} of connected pixels. The method consists in the following two steps (see Fig.~\ref{fig:DBSCAN}). 

%%%%%%%

\begin{enumerate}

%%%%%%%%
\item{{\bf Arcs and rings detection stage}}

\begin{itemize}
	\item An initial list of sources is extracted using the {\tt DBSCAN} algorithm (Top-left panel of Fig.~\ref{fig:DBSCAN}); 
	\item The shape of each source is determined and each source is flagged to be:
	arc-like, ring-like, ellipse-like, circular/point-like;
	\item All point-like sources are removed, leaving us with a list of candidate lensed sources;
	\item For each stamp in the candidate list, we compute the radial distribution of the sources, and we determine the minima and the maxima in this radial distribution. This is presented on the bottom-left panel of  Fig.~\ref{fig:DBSCAN}, where the  green filled circles  represent the local minima, and the red filled circles represent the local maxima. The first minima in the innermost ring sets the radius for the internal disk, indicated as a red circle in the top-left panel of  Fig.~\ref{fig:DBSCAN}. These allow us to partition the stamps in circular areas.
	\item All the sources within the internal disk (red circle in the left-top panel of Fig.~\ref{fig:DBSCAN}) are masked, i.e. all the corresponding pixels are set to the background flux level (central-top panel of Fig.~\ref{fig:DBSCAN});
	\item All the sources in the candidate list are assigned to a single circular area for each sources, enabling us to easily measure the angular size and the orientation of the sources, $\alpha_r$, with regard to the radial direction;
	\item All sources smaller than a critical angular size (e.g. the PSF size) are removed; only rings with at least one source meeting  criteria  are kept;
	\item The sources are then merged together in the final ring  to preform the analysis (see central panel in Fig.~\ref{fig:DBSCAN})
\end{itemize}
\vspace*{2mm}
\item{{\bf Analysis stage}}

\begin{itemize}
\item Once  we have a \textit{final} ring, we fit a circle to the  distribution 
of pixels in the recovered ring and we measure a centroid position, barycentre and 
radius, $R$ (green circle in the left panel of Fig.~\ref{fig:DBSCAN}). 
\item We also fit  a circle to each of the ring components that have an 
arc-like shape (blue dashed line in the right panel of Fig.~\ref{fig:DBSCAN}), 
and we check that the circle is contained within the final ring 
best-fit circle. 
\item We assign a quality factor to the ring, which is determined by the total angular 
coverage of the ring $\theta_{tot}$, and the displacement $d$,
between  the ring circle best fit centroid, and  the ring 
barycentre	
\begin{equation}
q_f=\frac{\theta_{tot}}{2\pi}\frac{1}{\exp(\frac{d-R}{R/f})+1}
\end{equation}
where the larger  the value of $f$, the more conservative  the quality factor is. In this work we adopt minimum $q_f=0.1$ to flag an object that are a possible lens.
\end{itemize}

\end{enumerate} 

We apply this automated procedure to the five bands of CFHTLS imaging data after PCA subtraction of the foreground object, leading to 1 098 lens candidates passing all above criteria.

\thispagestyle{empty}
\begin{table*}[t!]
\footnotesize
 \begin{center}
\begin{tabular}{ccccccccrc}

\hline\hline
ID  & RA        & DEC        & z & $R_{\rm eff}$   & $g$  & $r$   & $i$   &R$_{\rm E}$& Quality\\
\hline
  &         &         & & pix   & mag  & mag   & mag   &pix\\
\hline
      1&  30.2905&  -6.3474&  $0.548_{-0.507}^{+ 0.587}$&$   4.53$&$  21.28\pm   0.02$&$  20.21\pm   0.01$&$  19.61\pm   0.01$& 18.3&B\\
      2&  30.3615& -10.7597&  $0.563_{-0.521}^{+ 0.607}$&$   4.17$&$  23.37\pm   0.07$&$  21.96\pm   0.05$&$  21.17\pm   0.02$& 15.9&A\\
      3&  30.4522&  -7.5357&  $0.393_{-0.335}^{+ 0.437}$&$   4.51$&$  22.38\pm   0.04$&$  21.47\pm   0.03$&$  21.02\pm   0.02$&33.7&B \\
      4&  30.7655&  -4.4937&  $0.387_{-0.353}^{+ 0.423}$&$   3.90$&$  21.56\pm   0.02$&$  20.31\pm   0.01$&$  19.73\pm   0.01$& 12.1&B \\
      5&  30.9987&  -8.3652&  $0.495_{-0.459}^{+ 0.529}$&$   4.48$&$  21.48\pm   0.02$&$  20.30\pm   0.01$&$  19.76\pm   0.01$& 9.3&B\\
      6&  31.0361&  -9.6104&  $0.409_{-0.363}^{+ 0.463}$&$   3.83$&$  22.00\pm   0.02$&$  20.70\pm   0.01$&$  20.03\pm   0.01$& 16.9&A \\
      7&  31.2852&  -3.9099&  $0.252_{-0.227}^{+ 0.294}$&$   5.22$&$  20.56\pm   0.01$&$  19.76\pm   0.01$&$  19.31\pm   0.01$&7.06&A\\
      8&  31.4770&  -6.4598&  $0.442_{-0.411}^{+ 0.473}$&$   5.32$&$  20.98\pm   0.02$&$  19.69\pm   0.01$&$  19.17\pm   0.01$&7.8&B\\
      9&  31.9736&  -8.8432&  $0.289_{-0.243}^{+ 0.359}$&$   5.30$&$  20.55\pm   0.01$&$  19.83\pm   0.01$&$  19.47\pm   0.01$& 11.6&B\\
     10&  32.2221&  -6.9186&  $0.902_{-0.866}^{+ 0.939}$&$   6.40$&$  22.28\pm   0.04$&$  21.82\pm   0.05$&$  20.91\pm   0.03$&13.5 &A\\
     11&  32.3970&  -8.3013&  $0.611_{-0.580}^{+ 0.641}$&$   4.23$&$  21.85\pm   0.03$&$  20.63\pm   0.01$&$  19.81\pm   0.01$&8.6&A\\
     12&  32.4703&  -6.5295&  $0.389_{-0.351}^{+ 0.414}$&$   3.90$&$  20.00\pm   0.01$&$  19.12\pm   0.00$&$  18.65\pm   0.00$&9.03&A\\
     13&  32.5096&  -3.7956&  $0.556_{-0.515}^{+ 0.595}$&$   7.29$&$  21.60\pm   0.03$&$  20.48\pm   0.02$&$  19.86\pm   0.01$&9.4&A\\
     14&  32.6591&  -7.4773&  $0.478_{-0.453}^{+ 0.504}$&$   3.05$&$  22.93\pm   0.05$&$  21.44\pm   0.02$&$  20.64\pm   0.01$&30.0&A\\
     15&  32.8441&  -4.3681&  $0.731_{-0.703}^{+ 0.759}$&$   4.36$&$  23.13\pm   0.08$&$  21.81\pm   0.03$&$  20.49\pm   0.02$&17.4&A\\
     16&  32.9734&  -5.9950&  $0.139_{-0.050}^{+ 0.190}$&$   5.98$&$  19.29\pm   0.00$&$  18.69\pm   0.00$&$  18.31\pm   0.00$&6.6&B\\
     17&  33.0833&  -7.9352&  $0.451_{-0.428}^{+ 0.477}$&$   7.58$&$  21.06\pm   0.01$&$  20.26\pm   0.01$&$  19.64\pm   0.01$&6.7&A\\
     18&  33.1342&  -6.6479&  $0.450_{-0.421}^{+ 0.481}$&$   6.13$&$  20.94\pm   0.01$&$  20.01\pm   0.01$&$  19.63\pm   0.01$&9.2&A \\
     19&  33.1958&  -5.8338&  $0.809_{-0.768}^{+ 0.853}$&$   5.96$&$  21.76\pm   0.02$&$  21.16\pm   0.03$&$  20.36\pm   0.02$&6.3&B\\
     20&  33.6128&  -9.0673&  $0.698_{-0.670}^{+ 0.724}$&$   4.08$&$  22.44\pm   0.04$&$  21.25\pm   0.02$&$  20.17\pm   0.01$&8.3&A\\
     21&  33.6250&  -9.1754&  $0.398_{-0.366}^{+ 0.433}$&$   5.64$&$  21.16\pm   0.01$&$  20.01\pm   0.01$&$  19.59\pm   0.01$&10.1&A\\
     22&  33.8107&  -4.7156&  $0.346_{-0.313}^{+ 0.378}$&$   5.55$&$  20.40\pm   0.01$&$  19.32\pm   0.01$&$  18.83\pm   0.01$&10.3 &A\\
     23&  33.9600&  -4.4247&  $0.388_{-0.359}^{+ 0.417}$&$   5.23$&$  20.32\pm   0.01$&$  19.31\pm   0.01$&$  18.88\pm   0.01$& 9.4&B\\
     24&  34.9904&  -6.5704&  $0.486_{-0.464}^{+ 0.510}$&$   3.54$&$  22.72\pm   0.04$&$  21.26\pm   0.02$&$  20.43\pm   0.01$&13.8&A\\
     25&  35.0485&  -6.8143&  $0.489_{-0.454}^{+ 0.525}$&$   5.72$&$  20.85\pm   0.01$&$  19.90\pm   0.01$&$  19.45\pm   0.01$&6.5&A\\
     26&  35.0763&  -5.6397&  $0.709_{-0.676}^{+ 0.749}$&$   5.34$&$  22.29\pm   0.03$&$  21.66\pm   0.04$&$  21.09\pm   0.03$&6.2&B\\
     27&  35.1759&  -8.1834&  $0.361_{-0.337}^{+ 0.384}$&$   4.90$&$  20.51\pm   0.01$&$  19.53\pm   0.01$&$  19.18\pm   0.01$&7.2&A\\
     28&  35.3647&  -9.9535&  $0.772_{-0.732}^{+ 0.815}$&$   4.32$&$  23.06\pm   0.06$&$  22.19\pm   0.05$&$  21.13\pm   0.03$&21.9&A\\
     29&  35.5374&  -5.6453&  $0.457_{-0.420}^{+ 0.496}$&$   6.31$&$  21.06\pm   0.01$&$  20.21\pm   0.01$&$  19.93\pm   0.01$&7.5&A\\
     30&  37.0163&  -5.8651&  $0.400_{-0.356}^{+ 0.448}$&$   2.66$&$  22.78\pm   0.03$&$  21.68\pm   0.03$&$  21.27\pm   0.02$&7.6&B\\
     31&  37.1982&  -3.9803&  $0.680_{-0.651}^{+ 0.711}$&$   9.24$&$  21.66\pm   0.04$&$  20.70\pm   0.03$&$  19.93\pm   0.02$&9.8&B\\
     32&  37.5014&  -7.8604&  $0.591_{-0.549}^{+ 0.630}$&$   3.85$&$  22.96\pm   0.05$&$  21.79\pm   0.03$&$  21.01\pm   0.02$&12.7&B\\
     33&  37.5045&  -5.7003&  $0.559_{-0.525}^{+ 0.594}$&$   4.96$&$  21.31\pm   0.02$&$  20.16\pm   0.01$&$  19.54\pm   0.01$&7.6&A\\
     34&  38.0929&  -3.7355&  $0.798_{-0.762}^{+ 0.848}$&$   9.60$&$  22.03\pm   0.05$&$  21.38\pm   0.05$&$  20.41\pm   0.03$&10.5&B\\
     35&  38.2284&  -5.3160&  $0.710_{-0.665}^{+ 0.760}$&$   4.57$&$  23.10\pm   0.05$&$  22.36\pm   0.06$&$  21.73\pm   0.05$&10.7&A\\
     36& 132.1016&  -5.1126&  $0.412_{-0.388}^{+ 0.439}$&$   7.33$&$  21.02\pm   0.01$&$  19.99\pm   0.01$&$  19.50\pm   0.01$&7.1&A\\
     37& 132.1377&  -4.8329&  $0.774_{-0.745}^{+ 0.804}$&$   2.18$&$  22.99\pm   0.04$&$  21.86\pm   0.03$&$  20.65\pm   0.01$&13.3&A\\
     38& 132.5373&  -4.2216&  $0.222_{-0.186}^{+ 0.257}$&$   4.96$&$  19.66\pm   0.00$&$  18.70\pm   0.00$&$  18.25\pm   0.00$&9.8&A\\
     39& 133.2203&  -3.9328&  $0.430_{-0.401}^{+ 0.458}$&$   8.14$&$  20.49\pm   0.01$&$  19.49\pm   0.01$&$  19.12\pm   0.01$&10.6&A\\
     40& 133.5310&  -4.1230&  $0.670_{-0.641}^{+ 0.698}$&$   8.29$&$  22.18\pm   0.04$&$  21.22\pm   0.04$&$  20.44\pm   0.02$&14.5&A\\
     41& 133.7787&  -3.8646&  $0.715_{-0.687}^{+ 0.745}$&$   8.10$&$  21.60\pm   0.02$&$  20.67\pm   0.03$&$  19.85\pm   0.01$&10.4&A\\
     42& 134.4040&  -2.8884&  $0.687_{-0.660}^{+ 0.726}$&$   5.44$&$  22.05\pm   0.03$&$  21.37\pm   0.02$&$  20.76\pm   0.02$&7.2&B\\
     43& 135.0560&  -3.0676&  $0.608_{-0.566}^{+ 0.648}$&$   5.33$&$  22.53\pm   0.05$&$  21.33\pm   0.04$&$  20.54\pm   0.02$&9.1&A\\
     44& 135.2780&  -1.8642&  $0.353_{-0.322}^{+ 0.385}$&$   6.13$&$  20.47\pm   0.01$&$  19.51\pm   0.01$&$  19.11\pm   0.01$&11.2&A\\
     45& 135.4850&  -2.5134&  $0.689_{-0.663}^{+ 0.716}$&$   5.91$&$  22.16\pm   0.04$&$  21.22\pm   0.02$&$  20.49\pm   0.02$&6.7&A\\
     46& 208.9030&  57.0818&  $0.392_{-0.366}^{+ 0.420}$&$   5.15$&$  21.42\pm   0.02$&$  20.26\pm   0.01$&$  19.74\pm   0.01$&6.4&A\\
     47& 208.9420&  57.1261&  $0.406_{-0.383}^{+ 0.434}$&$   6.76$&$  20.32\pm   0.01$&$  19.33\pm   0.01$&$  18.92\pm   0.01$&11.0&A\\
     48& 209.2140&  54.2889&  $0.574_{-0.515}^{+ 0.616}$&$   5.09$&$  21.45\pm   0.02$&$  20.64\pm   0.01$&$  20.18\pm   0.01$&5.4&A\\
     49& 209.3525&  55.6741&  $0.398_{-0.374}^{+ 0.429}$&$   7.07$&$  20.29\pm   0.01$&$  19.31\pm   0.01$&$  18.93\pm   0.01$&14.2&B\\
     50& 209.3780&  53.4301&  $0.370_{-0.329}^{+ 0.399}$&$   4.96$&$  20.44\pm   0.01$&$  19.53\pm   0.01$&$  19.08\pm   0.01$&8.2&A\\
     51& 209.6380&  55.8449&  $0.376_{-0.352}^{+ 0.399}$&$   4.92$&$  20.02\pm   0.00$&$  18.99\pm   0.00$&$  18.55\pm   0.00$&9.8&A\\
     52& 209.7398&  57.0189&  $0.303_{-0.256}^{+ 0.343}$&$   6.29$&$  20.66\pm   0.01$&$  19.44\pm   0.01$&$  18.93\pm   0.01$&14.1&B\\
     53& 209.7620&  53.3673&  $0.331_{-0.277}^{+ 0.374}$&$   4.86$&$  21.10\pm   0.02$&$  19.77\pm   0.01$&$  19.21\pm   0.01$&12.2&A\\
     54& 209.8280&  57.4606&  $0.396_{-0.363}^{+ 0.432}$&$   5.92$&$  21.60\pm   0.02$&$  20.38\pm   0.01$&$  19.87\pm   0.01$&11.7&A\\
     55& 209.8940&  54.8880&  $0.456_{-0.424}^{+ 0.487}$&$   4.40$&$  20.61\pm   0.01$&$  19.81\pm   0.01$&$  19.45\pm   0.01$&7.1&B\\
     56& 209.8970&  56.7132&  $0.307_{-0.274}^{+ 0.339}$&$   5.00$&$  20.28\pm   0.01$&$  19.31\pm   0.00$&$  18.85\pm   0.01$&7.6&B\\
     57& 209.9210&  56.1383&  $0.370_{-0.340}^{+ 0.400}$&$   8.58$&$  20.52\pm   0.01$&$  19.41\pm   0.01$&$  18.92\pm   0.01$&11.2&A\\
     58& 210.0069&  56.9977&  $0.377_{-0.352}^{+ 0.398}$&$   7.27$&$  20.52\pm   0.01$&$  19.58\pm   0.01$&$  19.19\pm   0.01$&9.7&B\\
     \hline
\end{tabular}
\caption{List  of  grade-$A$ and grade-$B$ new lens candidates in CFHTLS. The photometric redshifts $z$, the effective radius $R_{\rm eff}$ and the magnitudes are the ones provided by \citet{Coupon:2009}. \label{GoodQuality1b}}
\end{center}

\end{table*}
\thispagestyle{empty}
\begin{table*}[t!]
\footnotesize
 \begin{center}
\begin{tabular}{ccccccccrc}

\hline\hline
ID  & RA        & DEC        & z & $R_{\rm eff}$   & $g$  & $r$   & $i$ &R$_{\rm E}$&Quality \\
\hline
&         &         & & pix   & mag  & mag   & mag   &pix\\
\hline
      59& 210.3022&  56.2394&  $0.412_{-0.382}^{+ 0.446}$&$  4.81$&$  20.93\pm   0.01$&$  19.58\pm   0.01$&$  18.99\pm   0.01$&14.7&B\\
     60& 210.3220&  57.3084&  $0.382_{-0.348}^{+ 0.412}$&$   7.61$&$  20.08\pm   0.01$&$  19.47\pm   0.01$&$  19.17\pm   0.01$&9.1&A\\
     61& 210.3420&  57.0673&  $0.810_{-0.757}^{+ 0.905}$&$   2.73$&$  22.92\pm   0.06$&$  22.52\pm   0.05$&$  21.98\pm   0.05$&7.9&B\\
     62& 210.5270&  53.4316&  $0.564_{-0.529}^{+ 0.601}$&$   7.80$&$  21.29\pm   0.03$&$  20.06\pm   0.01$&$  19.36\pm   0.01$&8.9&A\\
     63& 210.5496&  57.5600&  $0.759_{-0.727}^{+ 0.792}$&$   9.35$&$  21.89\pm   0.05$&$  20.97\pm   0.03$&$  20.11\pm   0.02$&14.8&A\\
     64& 210.5840&  51.7352&  $0.214_{-0.175}^{+ 0.250}$&$   4.40$&$  19.79\pm   0.00$&$  18.92\pm   0.00$&$  18.47\pm   0.00$&9.4&A\\
     65& 211.4080&  57.6165&  $0.306_{-0.271}^{+ 0.340}$&$   6.51$&$  20.33\pm   0.01$&$  19.43\pm   0.01$&$  19.03\pm   0.01$&8.9&A\\
     66& 211.8142&  57.1322&  $0.322_{-0.289}^{+ 0.353}$&$   6.09$&$  20.02\pm   0.01$&$  18.94\pm   0.00$&$  18.50\pm   0.00$&11.6&A\\
     67& 211.8690&  52.6938&  $0.485_{-0.448}^{+ 0.516}$&$   6.44$&$  21.18\pm   0.02$&$  20.03\pm   0.01$&$  19.56\pm   0.01$&7.1&A\\
     68& 211.9780&  56.2218&  $0.387_{-0.364}^{+ 0.412}$&$   7.21$&$  21.08\pm   0.01$&$  19.99\pm   0.01$&$  19.56\pm   0.01$&8.9&A\\
     69& 212.0175&  56.2446&  $0.370_{-0.339}^{+ 0.400}$&$   5.01$&$  21.05\pm   0.01$&$  19.88\pm   0.01$&$  19.36\pm   0.01$&10.1&A\\
     70& 212.1570&  52.3579&  $0.855_{-0.819}^{+ 0.899}$&$   6.39$&$  21.78\pm   0.03$&$  21.05\pm   0.03$&$  20.16\pm   0.02$&7.7&A\\
     71& 212.2455&  51.8158&  $0.343_{-0.315}^{+ 0.373}$&$   7.97$&$  21.02\pm   0.02$&$  20.55\pm   0.02$&$  20.50\pm   0.02$&11.2&A\\
     72& 212.3657&  53.5918&  $0.421_{-0.389}^{+ 0.453}$&$   5.84$&$  20.84\pm   0.01$&$  19.66\pm   0.01$&$  19.16\pm   0.01$&13.1&B\\
     73& 212.6040&  54.0908&  $0.420_{-0.383}^{+ 0.457}$&$   5.63$&$  20.84\pm   0.01$&$  19.68\pm   0.01$&$  19.25\pm   0.01$&12.4&A\\
     74& 212.7290&  54.9406&  $0.469_{-0.429}^{+ 0.510}$&$   4.91$&$  21.75\pm   0.02$&$  20.53\pm   0.01$&$  19.89\pm   0.01$&9.7&B\\
     75& 212.8450&  51.6687&  $0.499_{-0.455}^{+ 0.536}$&$   4.70$&$  21.66\pm   0.02$&$  20.41\pm   0.01$&$  19.80\pm   0.01$&6.5&A\\
     76& 213.0320&  52.9143&  $0.507_{-0.463}^{+ 0.560}$&$   6.89$&$  21.07\pm   0.02$&$  20.28\pm   0.01$&$  19.86\pm   0.02$&8.9&A\\
     77& 213.1650&  53.9570&  $0.436_{-0.403}^{+ 0.469}$&$   7.82$&$  20.64\pm   0.01$&$  19.53\pm   0.01$&$  19.05\pm   0.01$&8.2&A\\
     78& 213.4510&  51.7295&  $0.350_{-0.311}^{+ 0.384}$&$   7.57$&$  19.92\pm   0.01$&$  19.01\pm   0.01$&$  18.58\pm   0.01$&13.6&A\\
     79& 213.5430&  52.8470&  $0.393_{-0.366}^{+ 0.424}$&$   6.62$&$  20.86\pm   0.01$&$  19.83\pm   0.01$&$  19.40\pm   0.01$&7.3&A\\
     80& 213.6000&  57.6236&  $0.529_{-0.496}^{+ 0.559}$&$   5.81$&$  21.84\pm   0.03$&$  20.69\pm   0.02$&$  20.19\pm   0.01$&6.4&A\\
     81& 213.9140&  54.8451&  $0.232_{-0.202}^{+ 0.266}$&$   6.64$&$  18.79\pm   0.00$&$  18.05\pm   0.00$&$  17.65\pm   0.00$&11.9&A\\
     82& 214.4110&  56.3307&  $0.590_{-0.553}^{+ 0.619}$&$   5.54$&$  21.76\pm   0.02$&$  20.76\pm   0.02$&$  20.04\pm   0.01$&8.3&B\\
     83& 214.5100&  57.3730&  $0.570_{-0.535}^{+ 0.609}$&$   6.41$&$  21.15\pm   0.02$&$  20.19\pm   0.01$&$  19.74\pm   0.01$&7.7&A\\
     84& 214.5255&  54.2536&  $0.738_{-0.707}^{+ 0.775}$&$   4.95$&$  22.23\pm   0.03$&$  21.22\pm   0.03$&$  20.21\pm   0.01$&18.3&A\\
     85& 214.9620&  51.8585&  $0.682_{-0.658}^{+ 0.709}$&$   3.83$&$  22.01\pm   0.02$&$  21.16\pm   0.02$&$  20.45\pm   0.02$&8.9&B\\
     86& 215.3410&  56.2251&  $0.546_{-0.508}^{+ 0.583}$&$   5.86$&$  21.08\pm   0.01$&$  20.23\pm   0.01$&$  19.78\pm   0.01$&7.4&A\\
     87& 215.6690&  57.0355&  $0.433_{-0.401}^{+ 0.467}$&$   4.90$&$  20.54\pm   0.01$&$  19.41\pm   0.01$&$  18.91\pm   0.01$&9.2&A\\
     88& 216.3770&  56.4335&  $0.508_{-0.475}^{+ 0.538}$&$   7.41$&$  20.86\pm   0.01$&$  19.62\pm   0.01$&$  19.01\pm   0.01$&11.1&A\\
     89& 216.5700&  55.1213&  $0.629_{-0.602}^{+ 0.655}$&$   5.90$&$  21.73\pm   0.02$&$  20.66\pm   0.02$&$  19.80\pm   0.01$&9.2&A\\
     90& 216.7250&  56.1682&  $0.240_{-0.210}^{+ 0.268}$&$   6.96$&$  19.97\pm   0.01$&$  19.27\pm   0.01$&$  18.90\pm   0.01$&13.7&B\\
     91& 217.0550&  54.8198&  $0.855_{-0.814}^{+ 0.894}$&$   4.50$&$  23.40\pm   0.08$&$  22.12\pm   0.04$&$  20.97\pm   0.02$&18.4&B\\
     92& 217.1570&  55.4547&  $0.651_{-0.622}^{+ 0.679}$&$   6.09$&$  22.16\pm   0.03$&$  21.09\pm   0.02$&$  20.21\pm   0.02$&8.4&B\\
     93& 217.4450&  54.6213&  $0.639_{-0.609}^{+ 0.667}$&$   6.02$&$  21.55\pm   0.02$&$  20.84\pm   0.02$&$  20.28\pm   0.01$&6.5&B\\
     94& 217.9957&  55.7248&  $0.465_{-0.427}^{+ 0.507}$&$   8.14$&$  19.95\pm   0.01$&$  19.04\pm   0.00$&$  18.69\pm   0.01$&9.9&A\\
     95& 218.4500&  57.6522&  $0.307_{-0.274}^{+ 0.339}$&$   6.68$&$  19.74\pm   0.01$&$  18.70\pm   0.00$&$  18.26\pm   0.00$&11.8&A\\
     96& 218.9394&  55.9681&  $0.734_{-0.711}^{+ 0.758}$&$   7.15$&$  21.03\pm   0.01$&$  20.25\pm   0.01$&$  19.78\pm   0.01$&11.4&B\\
     97& 218.9660&  57.6901&  $0.616_{-0.563}^{+ 0.656}$&$   6.94$&$  21.96\pm   0.05$&$  20.99\pm   0.03$&$  20.26\pm   0.02$&12.4&B\\
     98& 330.2529&   2.2095&  $0.250_{-0.218}^{+ 0.285}$&$   5.42$&$  19.84\pm   0.01$&$  18.94\pm   0.01$&$  18.48\pm   0.01$&11.3&A\\
     99& 330.6014&   3.9024&  $0.316_{-0.284}^{+ 0.347}$&$   5.65$&$  20.17\pm   0.01$&$  19.03\pm   0.01$&$  18.56\pm   0.00$&11.8&A\\
    100& 330.6080&   2.1078&  $1.008_{-0.951}^{+ 1.062}$&$   6.13$&$  21.79\pm   0.03$&$  21.39\pm   0.04$&$  20.57\pm   0.03$&7.2&B\\
    101& 331.3547&   0.9742&  $0.621_{-0.588}^{+ 0.650}$&$   5.34$&$  22.00\pm   0.03$&$  20.95\pm   0.02$&$  20.26\pm   0.01$&8.3&B\\
    102& 331.6466&   2.2712&  $0.334_{-0.297}^{+ 0.375}$&$   3.88$&$  21.51\pm   0.02$&$  20.06\pm   0.01$&$  19.49\pm   0.01$&21.1&B\\
    103& 331.8250&   3.5431&  $0.411_{-0.380}^{+ 0.450}$&$   6.03$&$  21.45\pm   0.02$&$  20.40\pm   0.02$&$  20.00\pm   0.01$&7.9&B\\
    104& 331.8586&   1.4529&  $0.373_{-0.347}^{+ 0.396}$&$   7.66$&$  20.39\pm   0.01$&$  19.48\pm   0.01$&$  19.14\pm   0.01$&7.7&A\\
    105& 332.0030&   2.6561&  $0.466_{-0.426}^{+ 0.502}$&$   4.09$&$  21.76\pm   0.02$&$  20.64\pm   0.02$&$  20.15\pm   0.01$&10.7&B\\
    106& 332.3005&   3.7471&  $0.270_{-0.239}^{+ 0.298}$&$   4.36$&$  21.95\pm   0.02$&$  21.19\pm   0.02$&$  20.77\pm   0.02$&14.8&A\\
    107& 332.3815&  -0.2096&  $0.472_{-0.429}^{+ 0.509}$&$   8.11$&$  20.38\pm   0.01$&$  19.40\pm   0.01$&$  19.09\pm   0.01$&8.4&B\\
    108& 334.0200&   1.8810&  $0.764_{-0.735}^{+ 0.794}$&$   7.28$&$  21.75\pm   0.03$&$  20.98\pm   0.03$&$  20.14\pm   0.02$&11.2&A\\
    109& 335.5896&  -0.2775&  $0.291_{-0.257}^{+ 0.324}$&$   5.69$&$  20.14\pm   0.01$&$  19.12\pm   0.01$&$  18.62\pm   0.00$&13.3&B\\
\hline
\end{tabular}
\caption{List  of  grade-$A$ and grade-$B$ new lens candidates in CFHTLS. The photometric redshifts $z$, the effective radius $R_{\rm eff}$ and the magnitudes are the ones provided by \citet{Coupon:2009}. \label{GoodQuality1a}}
\end{center}
\end{table*}

%===============================
\section{CFHTLS results}
%===============================

In the following, we describe our main results using the PCA-finder. This includes a visualisation step by five of the authors, allowing us to define three subsamples of lens candidates depending on how likely the candidates are to be real lenses.  The characteristics of our new sample are compared with previous lens searches in the CFHTLS.
%==============================
%===============================
\subsection{Visual inspection}
%===============================

The automatically selected 1 098 candidates were  visually ins\-pected  to identify obvious spurious objects and to refine our lensing classification. We rank each object in one of the following categories: 
\begin{itemize}
\item ${\bf A}$: an almost  definite lens with a striking image configuration that is typical for lensing; 
\item ${\bf B}$: probable lens, but the candidate would need follow-up with spectroscopy or more imaging;
\item ${\bf C}$: possible lens, but with low probability of being confirmed, either because of low signal-to-noise (S/N) or because the potential lensed images are single or consist of short arcs that could still be compatible with edge-on galaxies or chain galaxies;
\item ${\bf 0}$: not a lens, spurious detection or spiral galaxy mimicking an arc or an Einstein ring. Objects in this category are false-positives. 
\end{itemize}
The visual classification is a time-consuming process. However, the workload remains reasonable in the case of the CFHTLS, which requires a few hours of human time to inspect the 1 098 candidates. The classification is performed both on the true-colour images using the $g$, $r$, and $i$ bands and on the residual images. 
This is done using the \textsc{FITS} images, enabling  us to easily and quickly explore the full dynamical range of the data. This classification is made independently by five of the authors: D.P., J.-P.K.,  R.J., F.C., P.D. 
 Out of all the systems, we select those objects that are classified as definite lenses by at least one individual initial judgment. All authors then needed to agree on a final classification.
Figures~\ref{Fig1}-\ref{Fig5} in the Appendix and Tables~\ref{GoodQuality1b}-\ref{GoodQuality1a} present all the lens candidates that we rank with the grade $A$ and grade $B$.

Our visual inspection shows that the most frequent conta\-minants are  face-on spiral galaxies, ring galaxies and polar ring galaxies. Face-on spirals mistakenly taken by the PCA-finder as lens candidates are easily identified by eye because their spiral arms point towards the bulge, while lensed arcs are tangentially aligned with respect to their lens galaxy. 

\begin{figure}[t!]
\centering
\hskip -8pt
\includegraphics[angle=0,width=.4\textwidth]{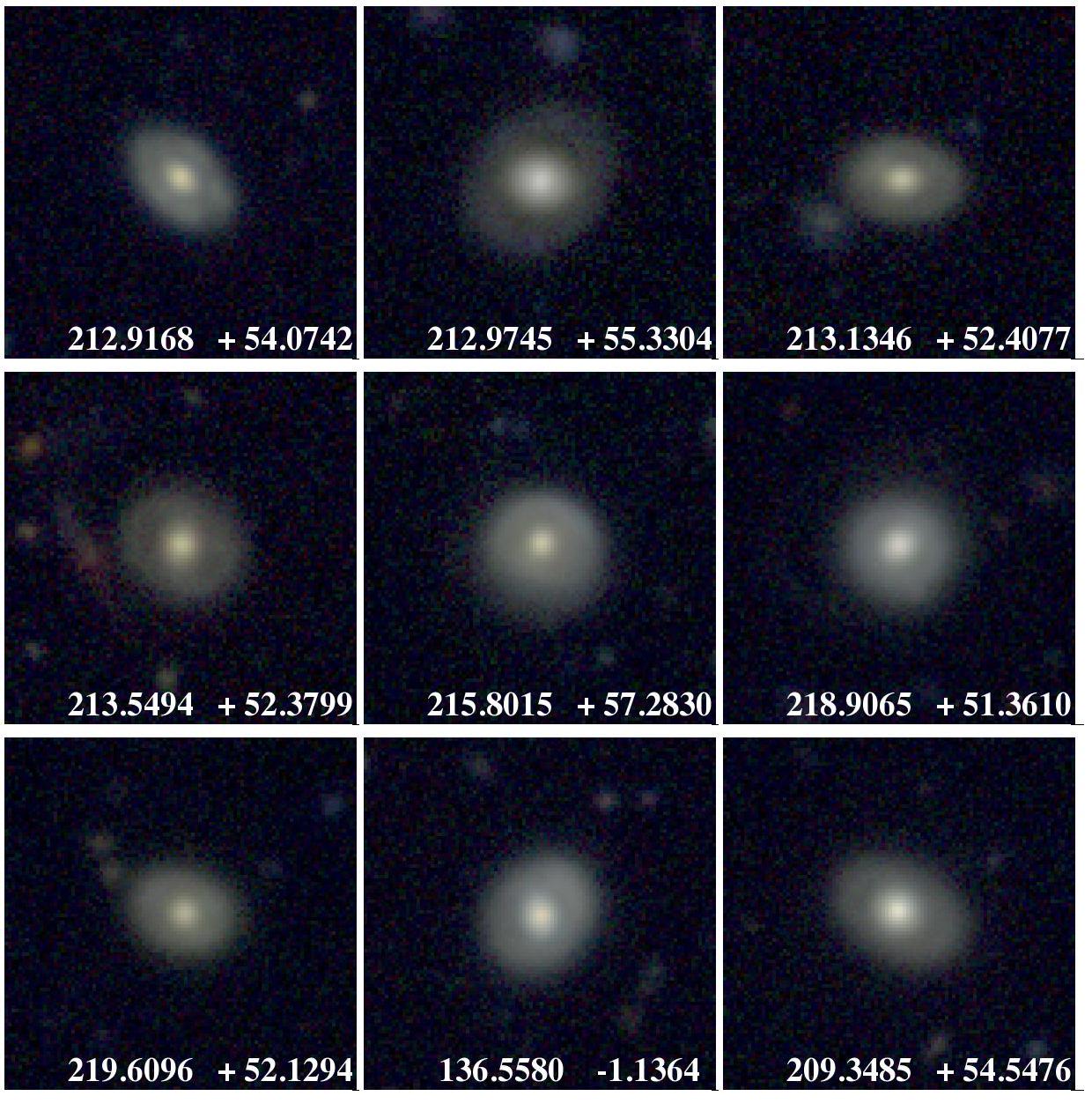} 
\caption{Examples of objects producing false positives in our lens search and that we classify as ring-like galaxies or polar ring galaxies (see text). Among 1 098 lens candidates, we identify 274 of these ring-like galaxies. }

\label{fig:polar}
\end{figure}

More problematic are ring-like galaxies in general and polar ring galaxies in particular. These rare composite galaxies  consist of a gas-poor, early-type galaxy (typically S0 galaxies) surrounded by a blue gaseous ring with ongoing star formation (see Fig.~\ref{fig:polar}). The most widely accepted explanation for the formation of polar ring galaxies is that accreted gas settles onto orbits that are more frequently contained either within the equatorial or  polar planes. Since the polar rings are blue and nearly perpendicular to the semi-major axis of their central hosts, they closely resemble Einstein rings that are produced by gravitational lensing. 

In our visual procedure, we attempt to classify an object as a ring galaxy if i) the ring structure has an ellipticity $\epsilon>0.2$, or ii) the ring shows a surface brightness close to constant. Our criterion on the ellipticity comes from the fact that only very extreme, rare, and rather unphysical lens galaxies or systems with extreme external shear can create a strongly elongated Einstein ring. Very elliptical Einstein rings are therefore not expected. In fact, none are known in the current literature. On the contrary, gaseous rings around polar ring galaxies can be strongly elliptical, simply due to orientation effects. The second condition, i.e. constant surface brightness, comes from the fact that Einstein rings are never fully symmetric and that lensed sources have structures, i.e. a bulge, spiral arms, etc. Ring galaxies have more uniform light distributions across the gaseous ring. 

Using the above criteria, the PCA-finder provides a list of 1 098 lens candidates that split, after visual classification, into:
\begin{itemize} 
\item 70 grade-$A$ candidates (Tables~\ref{GoodQuality1b} \& \ref{GoodQuality1a}),
\item 39 grade-$B$ candidates (Tables~\ref{GoodQuality1b} \& \ref{GoodQuality1a}),
\item 183 grade-$C$ candidates (Table~\ref{QualityB} of the Appendix),
\item 274 ring-galaxies or polar-ring galaxies (Table~\ref{PolarRing} of the Appendix).  
\end{itemize}

All our newly discovered grade-$A$ and grade-$B$ lens candidates are shown in Figs.~\ref{Fig1}- \ref{Fig5} of the Appendix \footnote{FITS images of our  lens candidates are also available  at \url{https://github.com/herjy/PiCARD}}. Despite the visual classification, all above candidates would need spectroscopic and/or high-resolution photometric follow-up, which is beyond the scope of this paper. In the following we compare the properties of our sample of 109 (new) grade-$A$ and grade-$B$ lenses with other lens samples found in the CFHTLS data. 

%===============================

\begin{table}[!th]
\footnotesize
\begin{center}
\begin{tabular}{rrrcl}
    \hline\hline
\multicolumn{5}{c}{\textsc{ArcFinder}  by  \citet{More:2015}} \\
  \hline\hline
  
  ID & RA & DEC & $i$ & $z_{\rm phot}$\\
  \hline
1   & 30.6619  & -6.5823   & 19.54 & 0.37 \\
2 & 30.8351  & -7.5808   & 19.45 & 0.59 \\
3 & 33.8459  & -7.6065   & 20.89 & 1.05 \\
4 & 35.2351  & -7.7199   & 20.51 & 0.71 \\
5 & 35.8142  & -6.4851   & 19.21 & 0.55 \\
6 & 36.5298  & -4.4573   & 17.97 & 0.17 \\
7 & 132.0986 & -4.1209  & 18.85 & 0.51 \\
8 & 134.4546 & -1.2169 & 18.26 & 0.29 \\
9 & 210.4371  & 53.0360   & 19.61 & 0.56 \\
10 & 214.8007  & 53.4365   & 19.11 & 0.69 \\
11 & 214.8219 & 51.2913   & 18.72 & 0.47 \\
12 & 217.9695  & 57.4769  & 20.19 & 0.83 \\
13 & 217.1451 & 52.2185   & 19.94 & 0.52 \\
14 & 217.5027 & 55.7799   & 19.12 & 0.55 \\
15 & 330.8709 & 2.0886   & 19.37 & 0.38 \\
16 & 331.2788 & 1.7844    & 19.15 & 0.46 \\
17 & 333.2789 & -0.5103   & 18.81 & 0.69 \\
18 & 333.5784 & 1.1761    & 18.84 & 0.74 \\
 \hline \hline
      \multicolumn{5}{c}{\textsc{ArcFinder} by  \citet{Maturi:2014} } \\
 \hline   \hline
1    &  33.5688    &  -5.0548 & 21.00&0.37\\
2    &  34.9856    &  -6.0341 &20.50&0.42\\
3    &  36.4030    &  -4.2549 &22.10&0.56\\
4    &  37.2865    &  -5.3320 &22.40&0.37\\
5    &  209.2597  & 52.5104 &23.00&0.38\\
6    &  209.6937  & 52.3495 &23.40&0.35\\
7   &  210.0883  & 52.2626 &21.20&0.76\\
8   &  335.5734  &   0.2007 &21.70&0.51\\
    \hline         
\end{tabular}
\end{center}
\caption{Strong lenses found using the two different  \textsc{ArcFinder}s (see text), and that we also find in the present work with the PCA-finder. \label{More-Maturi}}

\end{table}

\begin{table}
\footnotesize
\centering
\begin{tabular}{rrrcl}
  \hline\hline
\multicolumn{5}{c}{\textsc{RingFinder} by   \citet{Gavazzi:2014}} \\
  \hline\hline
      ID&RA&DEC& $i$ & $z_{\rm phot}$\\
  \hline
1& 31.0368 & -6.2019 & 19.92 & 0.440   \\
 2& 31.3527 & -9.5065  & 19.46 & 0.697   \\
 3& 32.7569 & -8.9320 & 20.67 & 0.562   \\
 4& 33.2527 & -8.7196 & 19.26 & 0.471   \\
 5& 33.9505 & -3.7979 & 19.7  & 0.577   \\
 6& 34.6119 & -7.2910 & 20.02 & 0.474   \\
 7& 35.2352 & -7.7199 & 20.48 & 0.688   \\
 8& 35.6735  & -5.6477 & 19.50  & 0.502   \\
 9& 36.4030 & -4.2549 & 19.60  & 0.631   \\
 10& 36.5152 & -9.7643 & 18.30  & 0.229   \\
 11& 36.6384 & -3.8179 & 20.08 & 0.652   \\
 12& 36.7455 & -8.0105 & 19.06 & 0.450   \\
 13& 37.1431 & -8.7207 & 19.08 & 0.493   \\
 14& 37.9618 & -4.2917  & 19.69 & 0.838   \\
 15& 38.6843 & -6.8091 & 20.32 & 0.728   \\
 16& 133.3229 & -2.0543 & 20.51 & 0.706   \\
 17& 133.7865  & -3.1020 & 20.64 & 0.613   \\
 18& 134.3794 & -1.0678 & 18.72 & 0.660   \\
 19& 136.5196 & -3.9364 & 19.51 & 0.776   \\
 20& 210.0897 & 51.5229  & 19.82 & 0.523   \\
 21& 210.0947 & 54.9680  & 20.01 & 0.703   \\
 22& 210.1774 & 56.0118  & 19.26 & 0.568   \\
 23& 210.6061 & 56.6629  & 20.32 & 0.662   \\
 24& 210.9173  & 56.7688  & 19.54 & 0.689   \\
 25& 211.0588 &51.7374 &19.69& 0.645 \\
 26& 211.1062 &52.0850& 18.82& 0.522  \\
 27& 211.3248 & 54.5971  & 20.48 & 0.726   \\
 28& 211.8857 & 54.5689   & 19.03 & 0.411   \\
 29& 212.2298 & 52.7479  & 19.98 & 0.492   \\
 30& 213.9302 & 52.4597  & 19.17 & 0.445   \\
 31& 214.8219 & 51.2913 & 18.72&  0.468  \\
 32& 215.0140 & 52.5271  & 20.99 & 0.510   \\
 33& 215.1154 & 54.1452  & 18.54 & 0.421   \\
 34& 215.1325 & 52.9728   & 18.87 & 0.461   \\
 35& 215.1830 & 54.8169  & 19.97 & 0.727   \\
 36& 215.8413 &  57.3786 &  19.11 &  0.611   \\
 37& 216.0988 & 52.5648  & 18.31 & 0.277   \\
 38& 216.1354 & 55.0055  & 19.39 & 0.451   \\
 39& 217.8799 & 57.1606  & 18.71 & 0.454   \\
 40& 218.5875 & 54.6375  & 20.29 & 0.728   \\
 41& 219.1557 & 54.9436  & 19.05 & 0.360   \\
 42& 219.7768 & 54.6502  & 19.33 & 0.378   \\
 43& 330.8709 & 2.0886  & 19.37 & 0.380   \\
 44& 331.2789 & 1.7845  & 19.15 & 0.460   \\
 45& 332.1596 & 3.0189  & 18.51 & 0.302   \\
 46& 333.3725 & 0.4932  & 18.91 & 0.483   \\
 47& 333.4006 & 0.1964  & 20.34 & 0.623   \\
 48& 333.4959 & 0.9046  & 18.27 & 0.370   \\
 49& 335.4535 & 1.2618  & 18.35 & 0.346   \\
 50& 335.5735 & 0.2008  & 19.13 & 0.421 \\
\hline

\end{tabular}
\caption{Strong lenses discovered with the \textsc{RingFinder} and also found by the PCA-finder. \label{ringfinder}}
\end{table}

\begin{table}[!th]
\footnotesize
\begin{center}
\begin{tabular}{rrrcl}
    \hline\hline
\multicolumn{5}{c}{\textsc{Space Warps} by   \citet{More:2015}} \\
  \hline\hline
  
    ID&RA&DEC& $i$ & $z_{\rm phot}$\\
  \hline
  1 & 31.6759 & -9.8669 & 20.8 & 0.2      \\
  2 & 32.1339  & -4.5542  & 21.0 & 1.0      \\
  3 & 33.1051 & -8.8697 & 19.5 & 0.8      \\
  4 & 135.5794 & -5.6566 & 0.0  & 0.0      \\
  5 & 211.5958 & 52.1617   & 20.3 & 0.7      \\
  6 & 216.5869 & 56.2323  & 19.5 & 0.5      \\
  7 & 216.7205 & 56.0016  & 0.0  & 0.0      \\
  8 & 217.3907 & 56.4277  & 19.0 & 0.5      \\
  9 & 217.7351 & 57.4084  & 19.3 & 0.7      \\
  10 & 219.2150 & 53.1183  & 19.2 & 0.6     \\
     \hline         
\end{tabular}
\end{center}
\caption{Galaxy-scale lensed systems found in the context of the \textsc{Space Warps} project that are also detected by our PCA-finder. \label{spacewarps}}

\end{table}

%===============================
\subsection{Comparison with previous searches}
%===============================

The CFHTLS data comprise all desirable survey properties for a lens search. They have been extensively explored in the past with a range of automated lens-finders, leading to very different lens samples. This clearly illustrates that no single technique can detect all the types of lenses at once and that current lens-finders are complementary. Some favour specific types of lensing object, such as spiral or elliptical galaxies, and others may select only massive lenses, e.g. by pre-selecting lenses as galaxy groups or clusters. Other favour a given source geometry, e.g. point sources (AGNs, quasars) or extended arcs and rings.  Here, we give a brief summary of previously published CFHTLS lens samples and we attempt to understand why the PCA-finder method finds some but not all lenses from the published samples.

A lens sample that is significantly different from the one in the present paper is provided by \citet{Elyiv:2013} and \citet{Sygnet:2010}. On the one hand \citet{Elyiv:2013} searched for gravitational lens candidates among the optical counterparts of X-ray-selected QSOs/AGNs. The authors visually inspect a sample of 5 500 optical counterparts of X-ray point-like sources identified in the XMM-LSS imaging of the CFHTLS W1 field. They find three good gravitational lens candidates. \citet{Sygnet:2010}, on other hand,  look for lensing events produced  only by massive edge-on disk galaxies. In their search, they preselect only highly elongated objects with $0.7>\epsilon>0.9$. Their final sample, which also involves a visual inspection, has 16 lens candidates. The PCA-finder neither looks for point-like multiple images nor for elongated lenses, thus we do not expect our search to recover any of those published lenses.

To the best of our knowledge, there are four lens searches similar to ours in CFHTLS \citep{Gavazzi:2014,More:2012,Maturi:2014,More:2015}. 
 \citet{More:2012} built a  sample of lenses using \textsc{ArcFinder} with a setting such that only systems with arc radii larger than 2\arcsec\ are kept in the sample. Their lens sample with large Einstein radii therefore predominantly selects group and cluster-scale lenses. \textsc{ArcFinder} measures the second order moments of the flux distribution in pixels within small regions of the sky to estimate the direction and extent of local elongation of features. Then, a set of thresholds on feature properties such as the area, length, width, curvature and surface brightness were used to select arc-like candidates. The search was carried out in the $g$ band which is the most efficient wavelength to find typical lensed features. The \textsc{ArcFinder} final sample consists of 55 promising lenses out of  a total of 127 lens candidates, which are selected from both CFHTLS {\it Wide} and {\it Deep} fields.  The PCA-finder recovers 16 out of these 127 candidates. This low fraction of recovered systems is somewhat expected since the majority of the systems found by \textsc{ArcFinder}  consist of  multiple lensing galaxies, which are not  recoverable by our method.  PCA-finder detects arcs and rings that are centred on single lensing galaxies, any of the lensing features around groups or cluster of galaxies are lost. Table~\ref{More-Maturi} lists the lenses found both by the PCA-finder and with \textsc{ArcFinder}.

%===============================

\citet{Maturi:2014} devised an automated lens-finder based on the colour statistics of arcs using a model for the spectral energy distribution (SED) of high redshift galaxies, the lensing optical depth, and the data depth. They therefore find lensed sources not only based on their morphology, but also from their colour, selecting the colours corresponding to sources providing the largest possible lensing cross-section.  Using this procedure, which combines the \textsc{ArcFinder}  created by \citet{Seidel:2007}, with a fine-tuned colour selection, they significantly increased the CFHTLS sample of gravitational lenses. They apply their method to the CFHTLS Archive Research Survey \citep[CARS;][]{Erben2009} data, which co\-vers 37 square degrees, to verify its efficiency and to detect new gravi\-tational arcs. Table~\ref{More-Maturi} lists the lenses found both by the PCA-finder and by \citet{Maturi:2014}.

 %===============================

\citet{Gavazzi:2014} use their \textsc{RingFinder} tool  to search for galaxies lensed by massive foreground early-type galaxies. The principle of \textsc{RingFinder} is similar to ours: they select all early-type galaxies from CFHTLS and then subtract them from the images to find lensing features.  There are two main differences between our work and \citet{Gavazzi:2014}: the way the lenses  are subtracted from the images and the way the residual images are analysed.  To remove the central galaxy,  \citet{Gavazzi:2014} subtracts the PSF-matched $i$-band images from the $g$-band images. On the residual image, they looked for excess flux in the $g$-band to search for compact lensing signal, i.e. multiply-imaged point sources, rings and arcs. In total 2 524 objects  passed their  automatic  selection procedure. These are visually inspected, leading to a total of 330 lens candidates, out of which 42 were ranked as good quality lenses and 288 were in their medium-quality category. In addition to the main  sample of \citet{Gavazzi:2014}, another 71 candidates were reported to have been detected by earlier versions of the \textsc{RingFinder}. From the main sample of \textsc{RingFinder}, during their follow-up campaign, they confirmed 33 lenses. Out of the 330 medium and high quality candidates found with \textsc{RingFinder}, 50 are also found by our PCA-finder (Table~\ref{ringfinder}).
 
Finally, the most recent CFHTLS lens search is known as \textsc{Space Warps} by \citet{More:2015} and is fully based on a visual detection and classification of lensing systems by humans, namely ``citizen" that volunteer to inspect the CFHTLS colour images. They report the discovery of 29 promising new lens candidates out of a total 59 candidates, based on about 11 million classifications performed by motivated citizen scientists. The goal of the blind lens search was to identify lens candidates missed by automated searches. This type of massive visual search  enables us to catch a larger diversity in lens and source properties than automated searches \citep[see also][for an example of a visual search in the HST database]{Pawase:2014}. Our PCA-finder recovers ten out of the 29 best candidates found in  \textsc{Space Warps}. These are listed in Table~\ref{spacewarps}.

Our PCA-finder cannot be expected to  recover all the \textsc{RingFinder} and \textsc{Space Warps} lenses, owing to the different levels of incompleteness of the different searches but also because of the pre-selection of galaxies in the PCA-finder search. We target early-type galaxies as potential lenses and we apply a cut in size for the lens galaxy (4-9 pixels). At least part of the lens candidates from the previous searches do not meet this size cut. We also note that the PCA-finder is not optimized to find  multiply-imaged point sources, which are very well spotted visually in \textsc{Space Warps} and with \textsc{RingFinder}.

Finding gravitational lenses is a  complex task, thus no single lens finding method is perfect, each method has  advantages over the other. It may be the case that a single method may never be the best method for optimising completeness and purity. As a consequence, it is not surprising that in spite of the many previous extensive lens searches in CFHTLS, we still manage to find new, interesting candidates. The PCA-finder, despite being very close conceptually to the \textsc{RingFinder}, has two major advantages. First, it is applied efficiently on single-band data. We therefore apply it independently to all the bands. In this way the search is not restricted to a limited range of source colours. Second, the resulting lens subtraction leaves very few artefacts, hence allowing us to spot fainter lensed features closer to the lens centre. 

We note that the 109 good lens candidates listed in Tables~\ref{GoodQuality1b} \& \ref{GoodQuality1a} are completely new. We also list  183 new objects in Table~\ref{QualityB} of the Appendix that we rank as possible lenses, but that certainly require follow-up with either deeper imaging or spectroscopy or both.

%===============================
\subsection{Sample properties}
%===============================

\begin{figure}[t!]
  \hspace{-1.2cm}
  \vspace{-4mm}
\includegraphics[scale=0.48,angle=0]{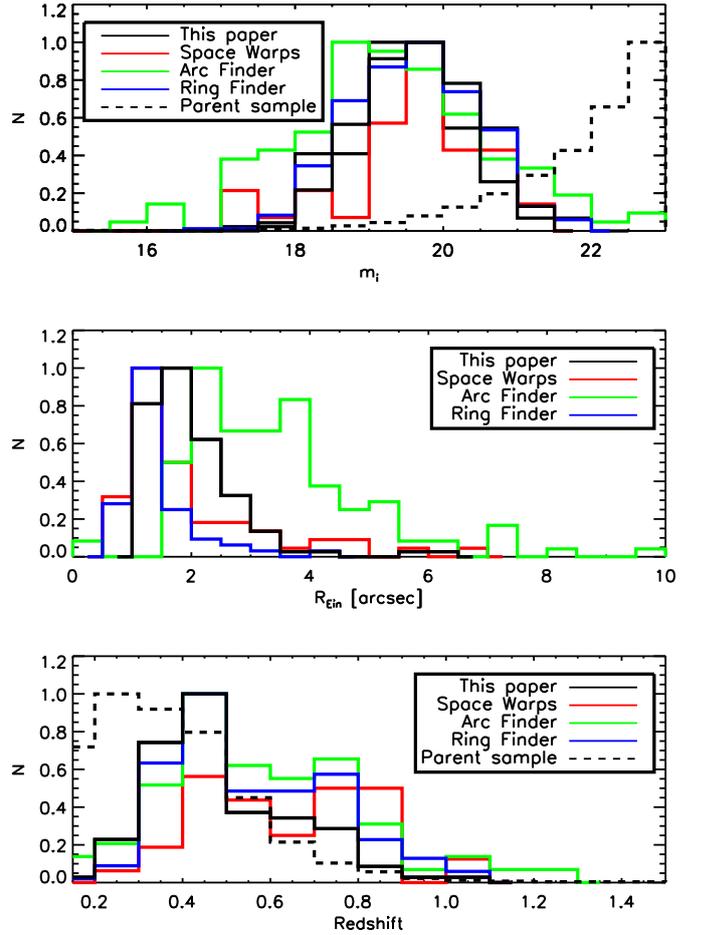}
\caption{$i$-band magnitude, Einstein radius, and redshift distributions of our lens candidates with $A$ and $B$ grades (black line). These are compared with the same distributions for other lens samples found in CFHTLS {\it Wide}: the one from the \textsc{Space Warps} program \citep[red line;][]{More:2015}, from the \textsc{ArcFinder} \citep[green line;][]{More:2012} and from the \textsc{RingFinder} \citep[blue line;][]{Gavazzi:2014}. When applicable, we also show the distribution for the parent sample, i.e. our preselection of potential lens galaxies.}
\label{histogram_properties}
\end{figure}

We now compare various properties of our lens candidates with previous samples from CFHTLS. We emphasise that these comparisons use lens candidates that are not yet confirmed  and that our results are therefore only indicative.

We use the CFHTLS photometric catalogues from \citet{Coupon:2009} and generated with the Le Phare software \citep{Ilbert:2006}. The accuracy of the photometric redshifts of galaxies for the {\it Wide} survey with magnitudes $i <21.5$ is $\sigma(\Delta z/(1+z)) = 0.032$.  Figure~\ref{histogram_properties}  shows the distributions in apparent magnitude, Einstein radii and redshift for our PCA-finder sample of 70 grade-$A$ plus 39 grade-$B$ new gravitational lens candidates. These are shown together with the same distributions for \textsc{Space Warps} \citep{More:2015}, for the \textsc{ArcFinder} \citep{More:2012}, and for the \textsc{RingFinder} \citep{Gavazzi:2014}.  We find that the median of the lens redshift distribution for the PCA sample is $z_{\rm PCA}=0.48 \pm 0.17$ and it is lower than redshift for all the other known lenses, which is $z_{\rm arcs}=0.52\pm0.20$ (including giant arcs which systematically have larger redshifts). The median of $i$-band magnitude of our sample is  $m_i=19.63$, which turns into a median absolute magnitude of our sample $M_{g}=-21.90\pm0.745$. These magnitudes are K-corrected following \citet{Coupon:2009}.
     
The Einstein radii displayed in Fig.~\ref{histogram_properties} are estimated from the position of the multiply-lensed images. $R_E$ is taken to be half the averaged values of the  angular separation between  images. The distribution of image separations  can be used to probe the average density profile of the lens population \citep{Oguri:2006,More:2012}. We find that the average  Einstein radius for our new lenses is $R_E=1.9\pm0.8\arcsec$ which is, as expected, smaller than for the \textsc{ArcFinder} candidates, which have $R_E=4.0$\arcsec. This is also lower than for \textsc{Space Wraps} which have $R_E=1.9$\arcsec. For comparison the \textsc{SLACS} lenses have $R_E=2.2\arcsec$ and the \textsc{RingFinder} lenses is $R_E=1.4\arcsec$ \citep[see][]{Sonnenfeld:2013}.

\section{Simulations and completeness}

An evaluation of the completeness of our sample can be done in two ways,  using realistic image simulations or using a sample of already known lenses. The latter approach has been attempted in the previous section, but has a clear limitation: the reference sample of known lenses has its own completeness and purity.   Moreover, different lens-finders are not necessarily optimised to find the same lens types and the parent samples (i.e. after the pre-selection) are not the same. As  was shown in the previous section, one lens-finder can be very efficient at finding objects with low lens/source luminosity contrast, another one may  be specialised in finding large arc-like structures and others may find better lensed point sources. For all these reasons, we choose to use simulated images for our completeness estimation. 

In this section we evaluate the performances of our method using simulated images of Einstein rings,  as they would be seen with the CFHT. We have made an attempt to generate lenses that are as  realistic  as possible within the requirements of the PCA lens-finder.  In the following, we also describe some of the properties of our simulated sample.

\begin{figure*}[t!]
\centering
  \begin{tabular}{cc}
\includegraphics[width=.43\textwidth]{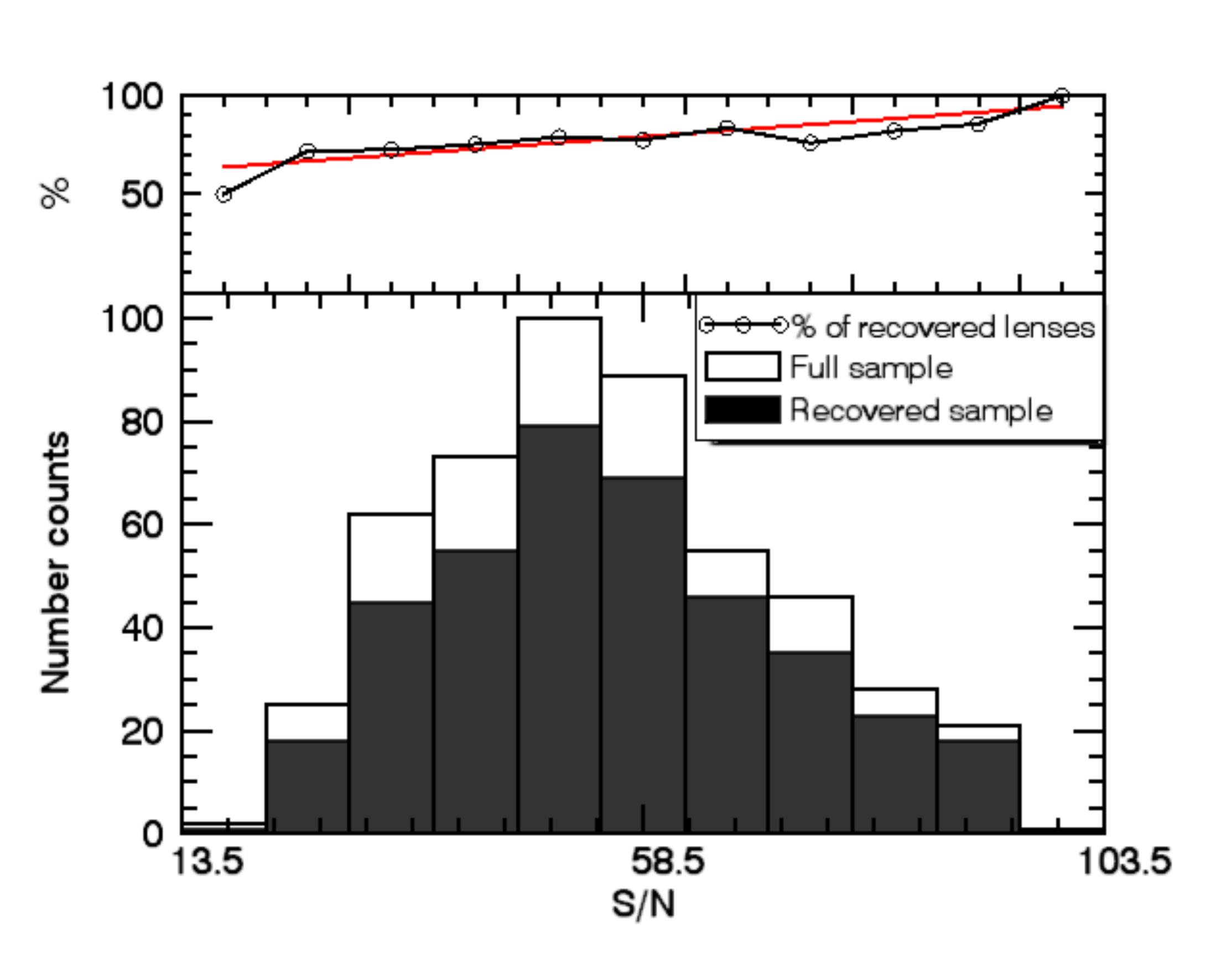}  &
\includegraphics[width=.43\textwidth]{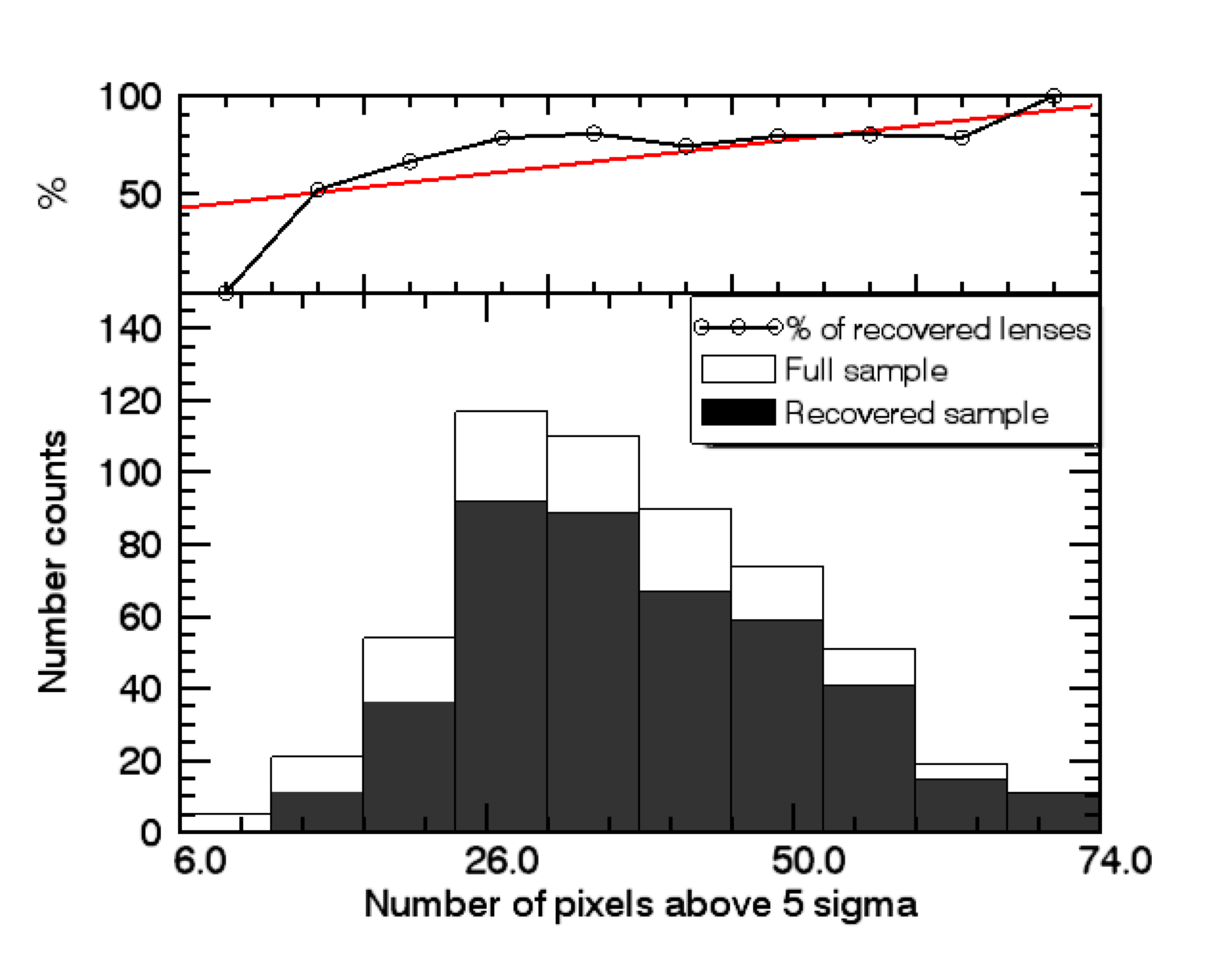} \\
\includegraphics[width=.43\textwidth]{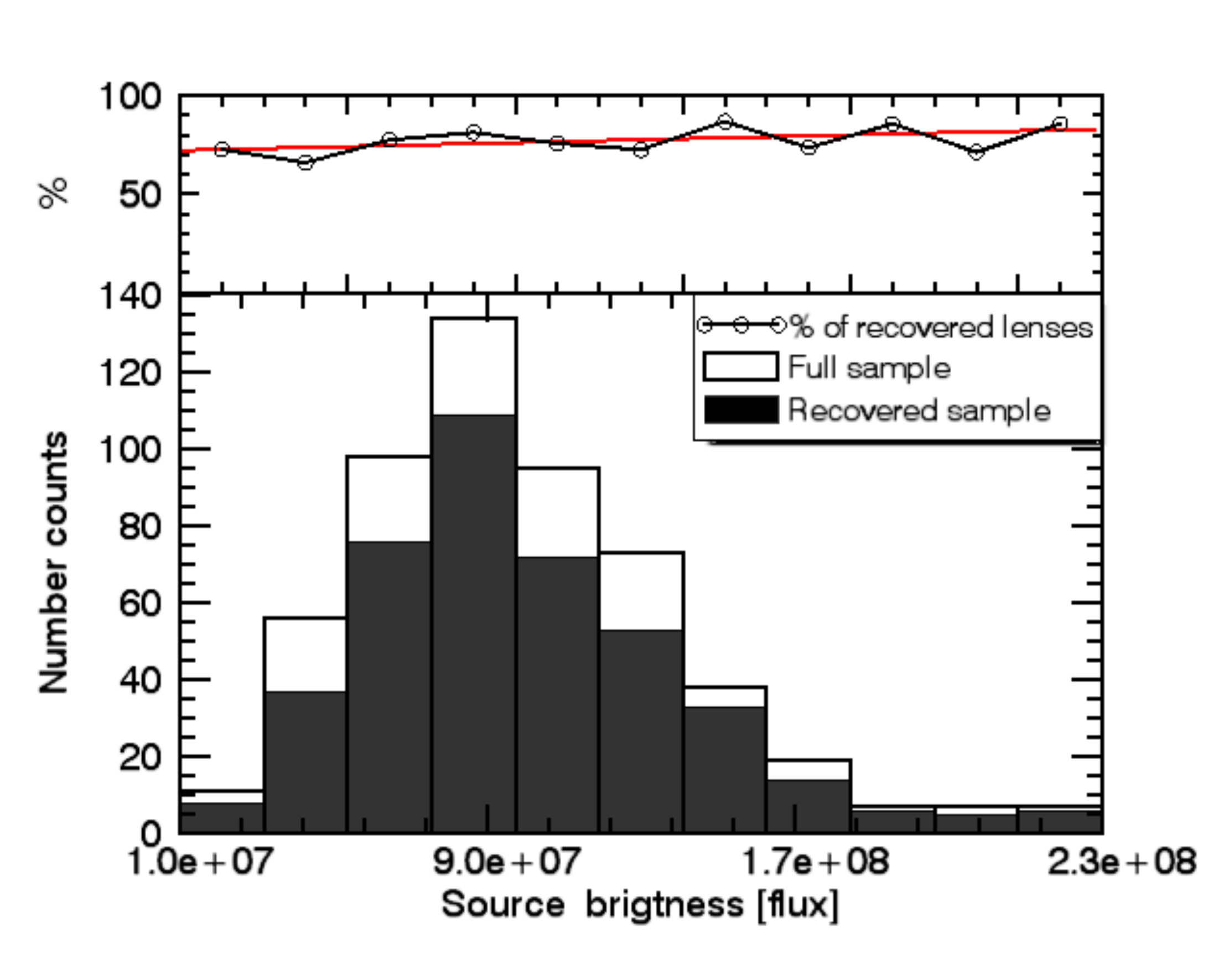}  &
\includegraphics[width=.43\textwidth]{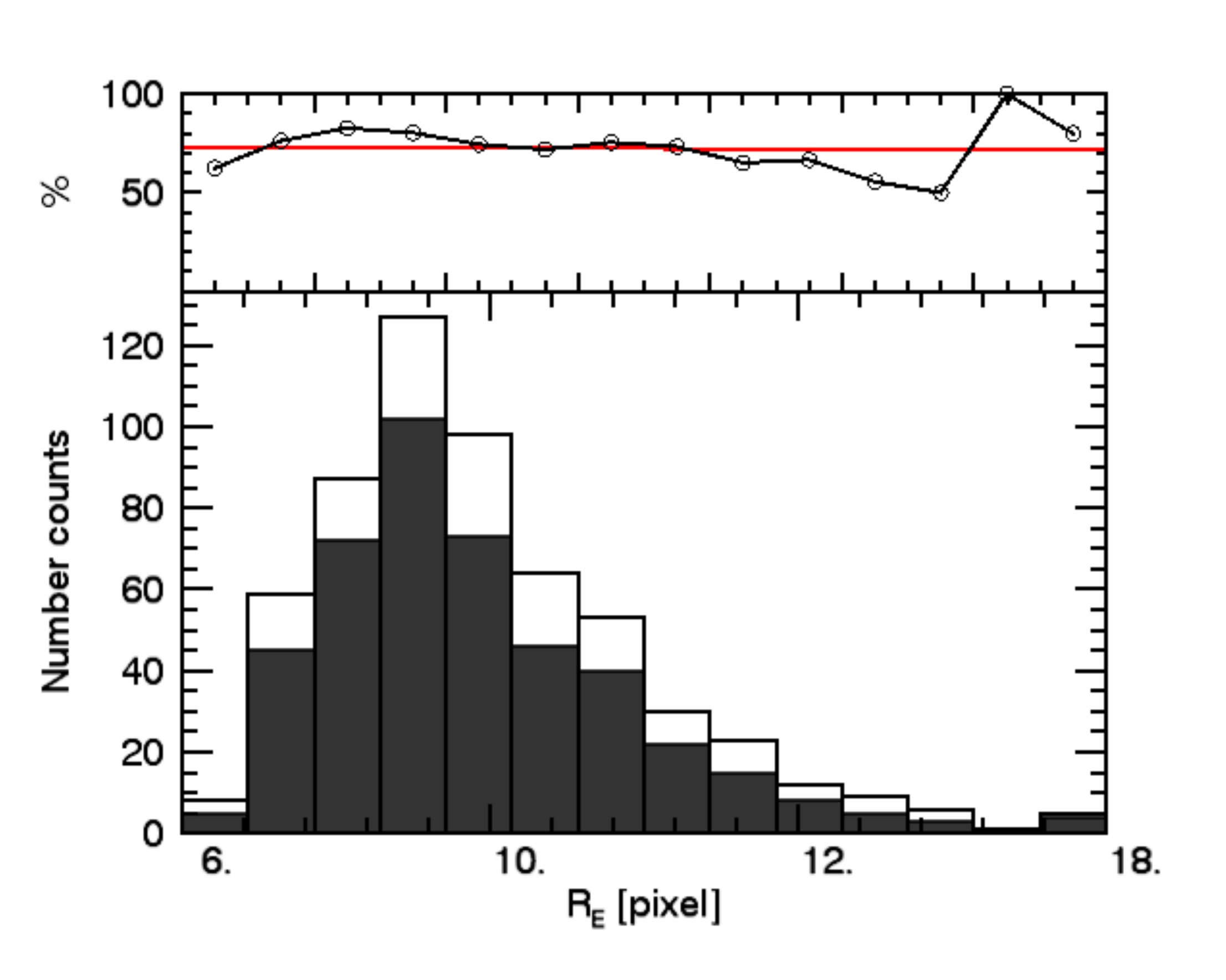}  \\
\includegraphics[width=.43\textwidth]{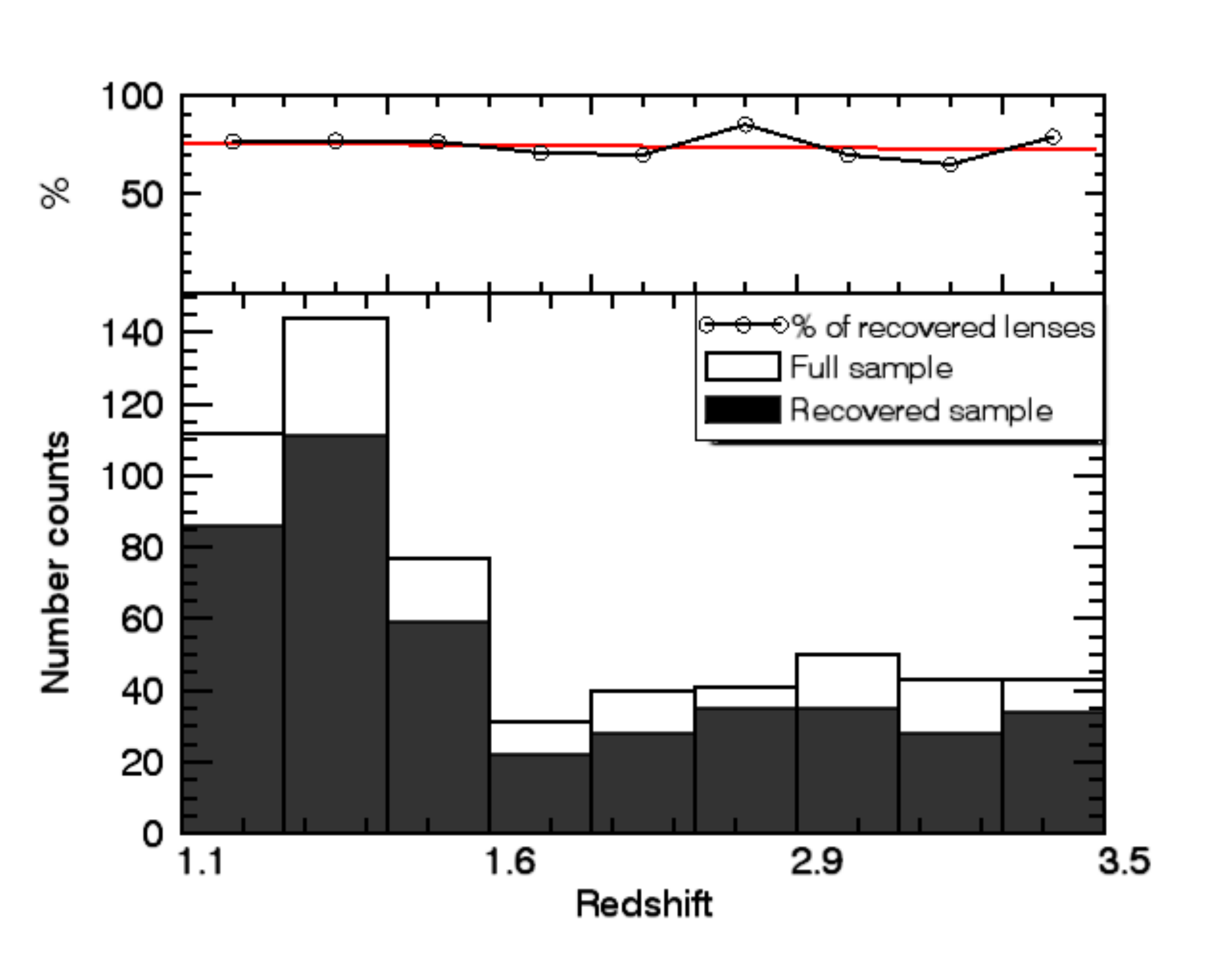}  
\end{tabular}
\caption{Properties of the lensed systems for the simulated sample. Each plot in the lower panels shows the distributions of selected parameters of the full sample of simulated systems and for the simulated objects we  identify as lenses (true positives). The top panels give the ratio of the two, i.e. the completeness per bin of the selected parameter. The red line is a linear regression to guide the eye. }
\label{simulation}
\end{figure*}

%===============================
\subsection{The lens and source simulation}
%===============================

The image simulations are provided by the Bologna Lens Factory (BLF).
The BLF setup was chosen to match the properties of gravitational lenses expected in the CFHTLS wide fields by adding fake lensed objects to real images.

 The lensing simulations were done as follows. A dark matter halo catalogue within a light cone extending out to $z=4$ is taken from the Millennium Run Observatory \citep{2013MNRAS.428..778O}.  This contains 
all the halos found within the Millennium cosmological simulation that were more massive than $10^{10}~M_{\odot}$, which should include the hosts of all the observable strong lenses.  The light cone covers 1.6 square degrees of sky. 
 Each halo is represented within the lensing code by a Navarro, Frenk \& White (NFW) halo plus a singular isothermal ellipsoid (SIE) in its centre to represent the baryonic component. 
 The mass of the baryonic component  is determined using the halo mass vs. stellar mass relation of \cite{2010ApJ...710..903M}  and 
the velocity dispersion is set by the Tully-Fisher relation \citep{2001ApJ...550..212B}.  Once the light cone is assembled, all the caustics in this light cone with Einstein radii $R_E >0\farcs05$  are located, for a series of six source planes running from $z=1$ to $z=3.5$, using the GLAMER lensing code \citep[][]{GLAMERI, GLAMERII}.  GLAMER shoots rays through the light cone and identifies regions of the source plane that will be strongly lensed.  The code then adaptively shoots rays in these areas at higher and higher resolution to resolve the critical curves and caustic curves of each prospective lens.

The simulated image of a  lensed source  is then added to a real image of a galaxy that is randomly taken from the preselected
CFTHLS data.  This step was unavoidable, bearing in mind that the PCA lens extraction is solely based on the self-similarity of the foreground galaxies. Since the simulations  need to be as close to real CFHTLS
data as possible, we thus draw galaxies from the preselected targets in size, magnitude, colour, etc (see Sect.~\ref{selection}). Galaxies selected for the simulations in such manner,  contain all the relevant
limits that we face in real data. This gives us several advantages: 1. the simulation naturally includes the noise properties of the original data, 2. the level of complexity in galaxy shapes is well
 representative of the real data, far beyond the reach of analytical galaxies models, 3. the simulation includes blending effects with companion galaxies. Even though this approach gives us a reliable way to
estimate the completeness of our sample,  it does not allow us to evaluate its purity. This requires  a priori knowledge of which galaxies are and which  are not acting as lenses. Since the galaxies
are drawn randomly from the real data, they can potentially contain lensed features that would affect the results. While a visual inspection would probably solve this problem, it could however bias the completeness.

We note that the distribution of lens properties should be generally reproduced in  the simulations, but that it is not necessary to reproduce the statistics of the lenses 
to high precision for our purposes.  In the next section, we  characterise the lens-detection efficiency in terms of various parameters such as S/N, Einstein radius, etc.   
It is necessary that the simulations fully cover the range of these parameters, but not that they reproduce the predicted distribution of parameters precisely.  These simulations meet this requirement.

%===============================
\subsection{Completeness of the new lens sample}
%===============================

Our simulations include 600 systems, which match well the properties of the galaxies we preselected in CFHTLS (Sect.~\ref{selection}), as well as the noise properties of the images. We run the PCA lens-finder in the exact same way as we do on the real data, exclu\-ding the last step of a visual inspection. Our results are summarised in Fig.~\ref{simulation}, where we compare the distribution of some of the most important observational parameters for the  full population of the  simulated lenses and for the population of simulated systems, actually identified as such by the PCA lens-finder. The ratio between the two histograms in Fig.~\ref{simulation} gives the completeness.

In our analysis, we estimate the completeness as a function of the total S/N in the (lensed) source, the number of pixels of the source above the noise level (5$\sigma$), the source surface brightness, $R_E$  (taken as half the averaged values of the  angular separation between  images) and the source redshift. We note that the calculation of the source S/N includes a noise contribution from the lensing galaxy, which can significantly impact the detection when the source and the lens overlap. Using our lens simulations we find that:

\begin{itemize}

\item Not surprisingly the completeness of the sample improves with increasing S/N, reaching at least 80\% as soon as $S/N>50$. Even for the lowest source S/N in the sample, the completeness is still above 50\% and this number increases to 70\% when $S/N>20$. Of course, within a given S/N bin the sample spans a large range of Einstein ring size and source/lens luminosity contrast, but overall the completeness achieved by the PCA lens-finder is very high.

\item The completeness depends strongly on the number of pixels above a given luminosity threshold of the lensed source. This affects the angular size of the lensed image and therefore also our ability to determine its shape (ring, full, or partial arc). Of course the more spatial resolution elements in a lensed source, the better it can be classified.

\item The completeness has slightly weaker dependence on source surface brightness than on S/N of the lensed source.  This shows that we fail to  detect some lenses with arcs and rings that fall  into  the glare of the lensing galaxy. This also shows that central lensing galaxy  removal with PCA method is not perfect and has an impact on the lens  search.  
 
\item Importantly, there is little or no dependence of the completeness with respect to $R_E$ or to the source redshift. This suggests that the PCA lens-finder is capable of providing unbiased samples of systems spanning a broad range of masses and redshifts, which is desirable for galaxy formation and evolution studies based on strong gravitational lensing.

\end{itemize}

\section{Summary and conclusions}

We have implemented a novel method, PCA-finder, for the automated detection of galaxy-scale strong gravitational lensing  to a heavily explored survey, namely,  the 155 square degree of imaging data of the CFHTLS {\it Wide}. With the PCA lens-finder we discover 109 (70 grade $A$ and 39 grade $B$) brand new gravitational lens candidates.    
The discovery of such a large number of  new lens candidates missed  by other searches  proves  PCA-finder to be a powerful tool in discovering lenses.

The search was carried out in four steps. In Step 1, we create a uniform data cube consisting of small image stamps  centred on preselected early-type  galaxies. 
In Step 2, we subtract central galaxies  from the image stamps. In Step 3, we analyse residual images created in Step 2 to look for lensing features with the \texttt{DBSCAN} method. In Step 4, a  sample of 1 098 candidates are selected from this automated procedure.  Finally, Step 5 is a visual inspection of the lens candidates by five  authors of this paper (D.P., J-P.K.,  R.J., F.C., P.D.). Following this last step, all candidates are  allocated $A$, $B$, or $C$ grades.

In this paper, we present the new PCA-finder lenses and compare it with the previously known samples from the CFHTLS, namely, \textsc{Space Wraps}, \textsc{RingFinder} and \textsc{ArcFinder}.
Our main results can be summarised as follows:
\begin{itemize}
\item PCA-finder works well as a discovery engine for gravitational lenses.
\item We present a sample of 70 grade-$A$ and 39 grade-$B$ new gravitational lens candidates, and  additional 183 grade-$C$  worth noticing, but with no strong evidence for lensing. We rediscover 86 lens candidates
from various samples published in the literature.
\item The PCA-finder selects lens systems  whose statistical properties are well comparable with the  \textsc{RingFinder} and \textsc{ArcFinder} samples, including the range of lens redshifts, magnitudes, and image separation.
\item We also find  274 potential ring galaxies or polar ring galaxies.
\item  We use a sample of simulated lenses tailored to the CFHTLS {\it Wide} data to verify the completeness of our automated method.
\end{itemize}

The discovery of many new lens candidates through the first PCA-finder lens search illustrates the strength of the method, since we find lens candidates that other algorithms missed. Upcoming and planned wide field imaging surveys such as the DES, HSC, KIDS, Euclid and the LSST will produce a great amount of data. Reliable automated algorithms  together with citizen blind search will be necessary to  find lenses in these  very large surveys. As shown in this paper, one approach  for  finding lenses from the entire survey data may not be sufficiently complete and pure. Thus, combining robotic methods for pre-selection with the citizen science approach for visual screening might be a good strategy for finding lenses in these large imaging surveys. For samples that are sufficiently cleaned by the automated part of the pipeline, the human time spent on the final classification remains acceptable.

\acknowledgements{This research is supported by the Swiss National Science Foundation (SNSF). B. Metcalf's research is funded under the European Seventh Framework Programme, Ideas, Grant no. 259349 (GLENCO).}

\bibliographystyle{aa}
\bibliography{DanutaParaficz_arxiv}

\begin{appendix}
\section{Observational data}

\begin{figure*}
\centering
\includegraphics[width=.8\textwidth]{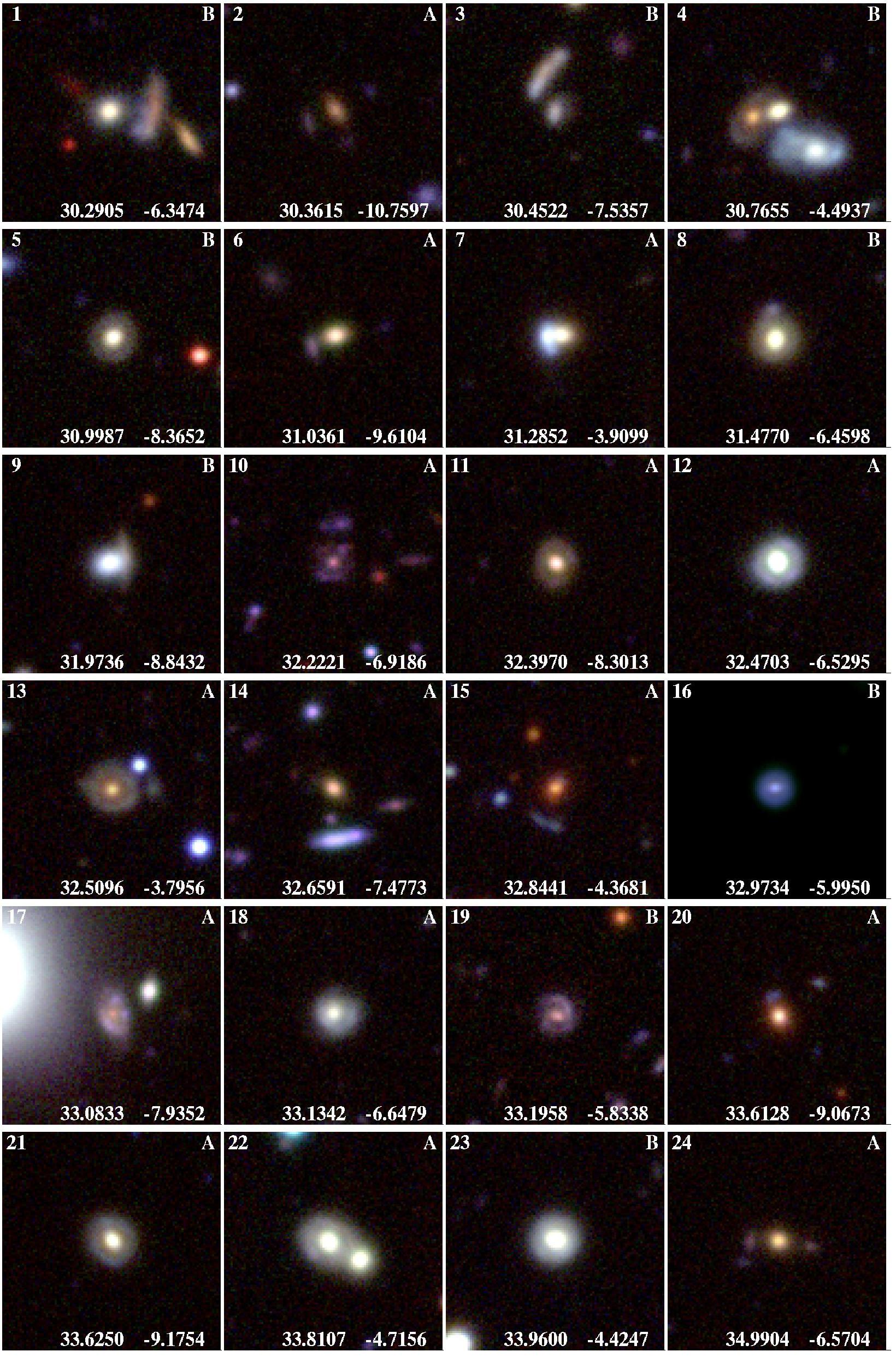}  
\caption{Our new lenses found with the PCA-finder, with grades $A$ and $B$ (see text). The stamps are 18.7\arcsec on-a-side. \label{Fig1}}
\end{figure*}

\begin{figure*}

\centering

\includegraphics[width=.8\textwidth]{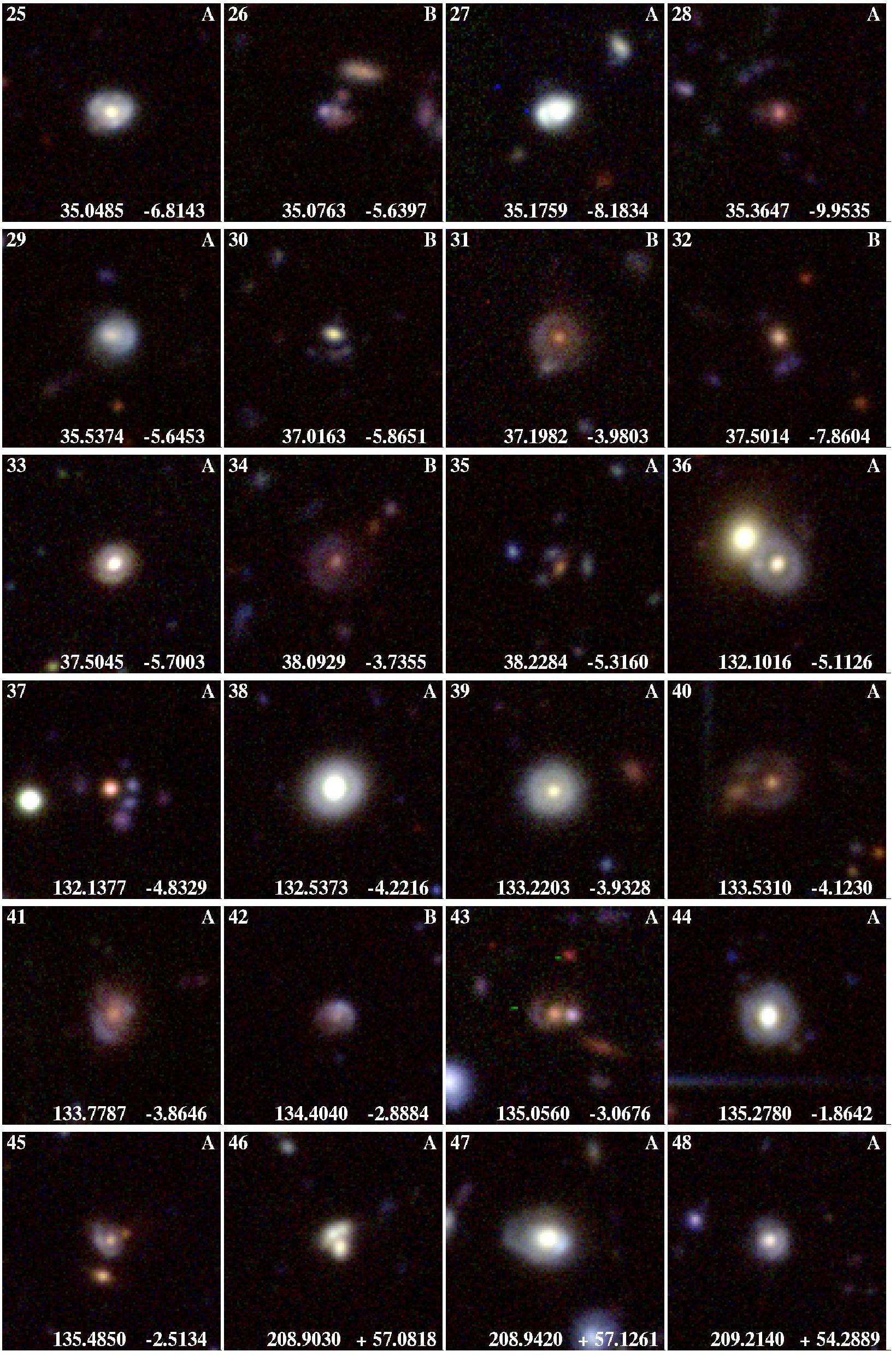}

 \caption{Our new lenses found with the PCA-finder, with grades $A$ and $B$ (see text). The stamps are 18.7\arcsec on-a-side. \label{Fig2}}
\end{figure*}

\begin{figure*}
\centering
\includegraphics[width=.8\textwidth]{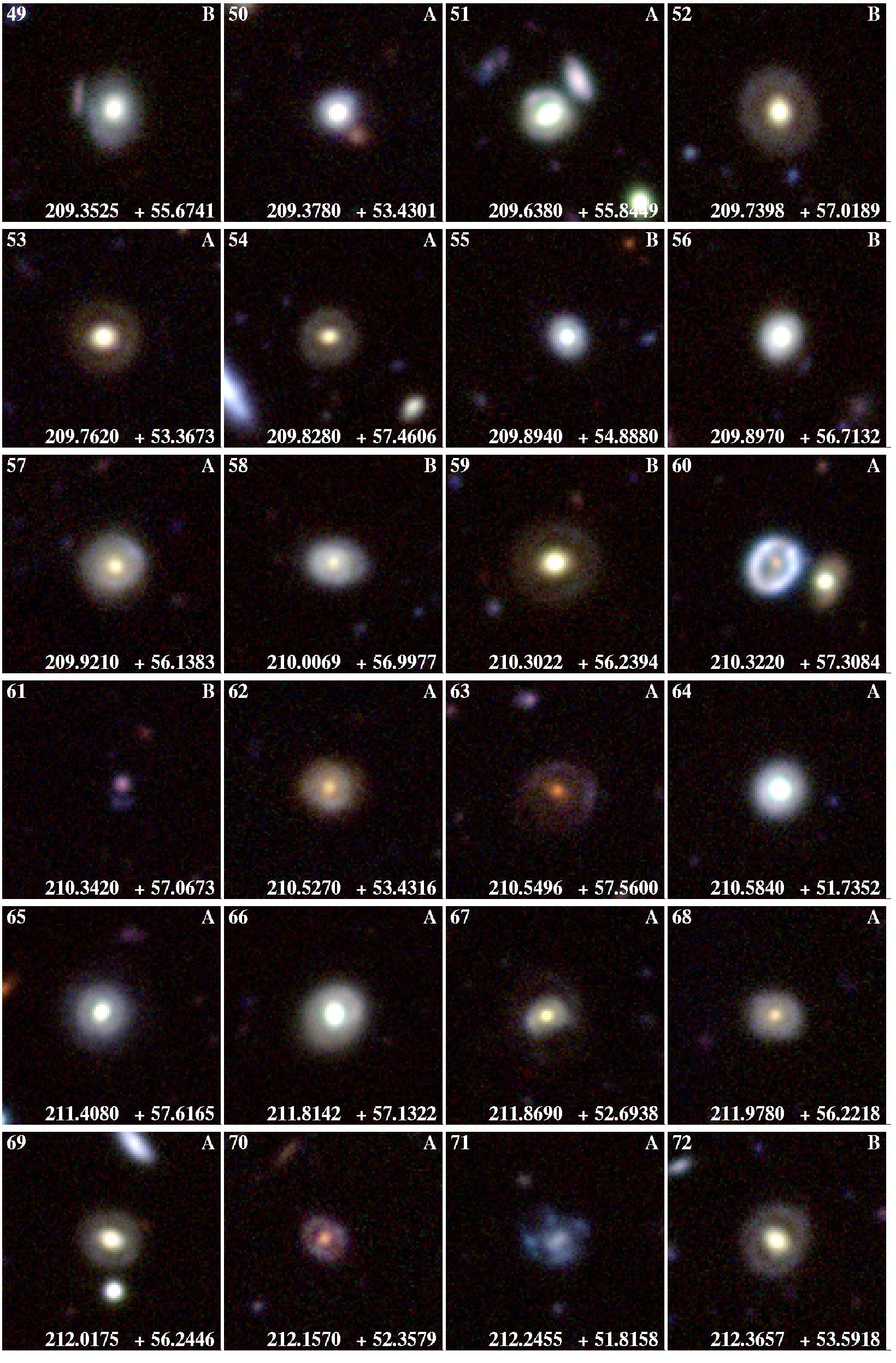}
\caption{Our new lenses found with the PCA-finder, with grades $A$ and $B$ (see text). The stamps are 18.7\arcsec on-a-side. \label{Fig3}}
\end{figure*}

\begin{figure*}
\centering
\includegraphics[width=.8\textwidth]{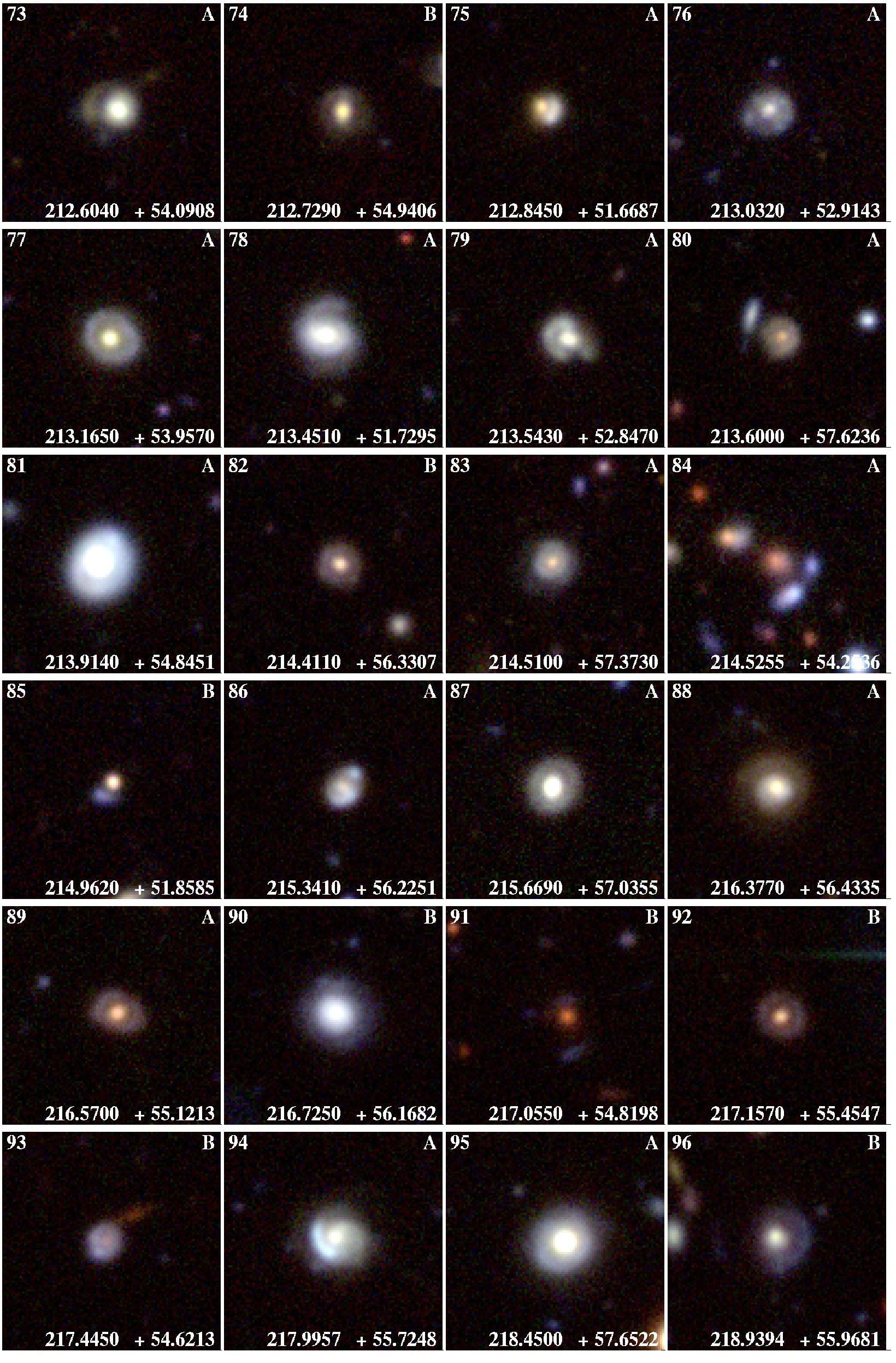}
\caption{Our new lenses found with the PCA-finder, with grades $A$ and $B$ (see text). The stamps are 18.7\arcsec on-a-side. \label{Fig4}}
\end{figure*}

\begin{figure*}
\centering
\includegraphics[width=.8\textwidth]{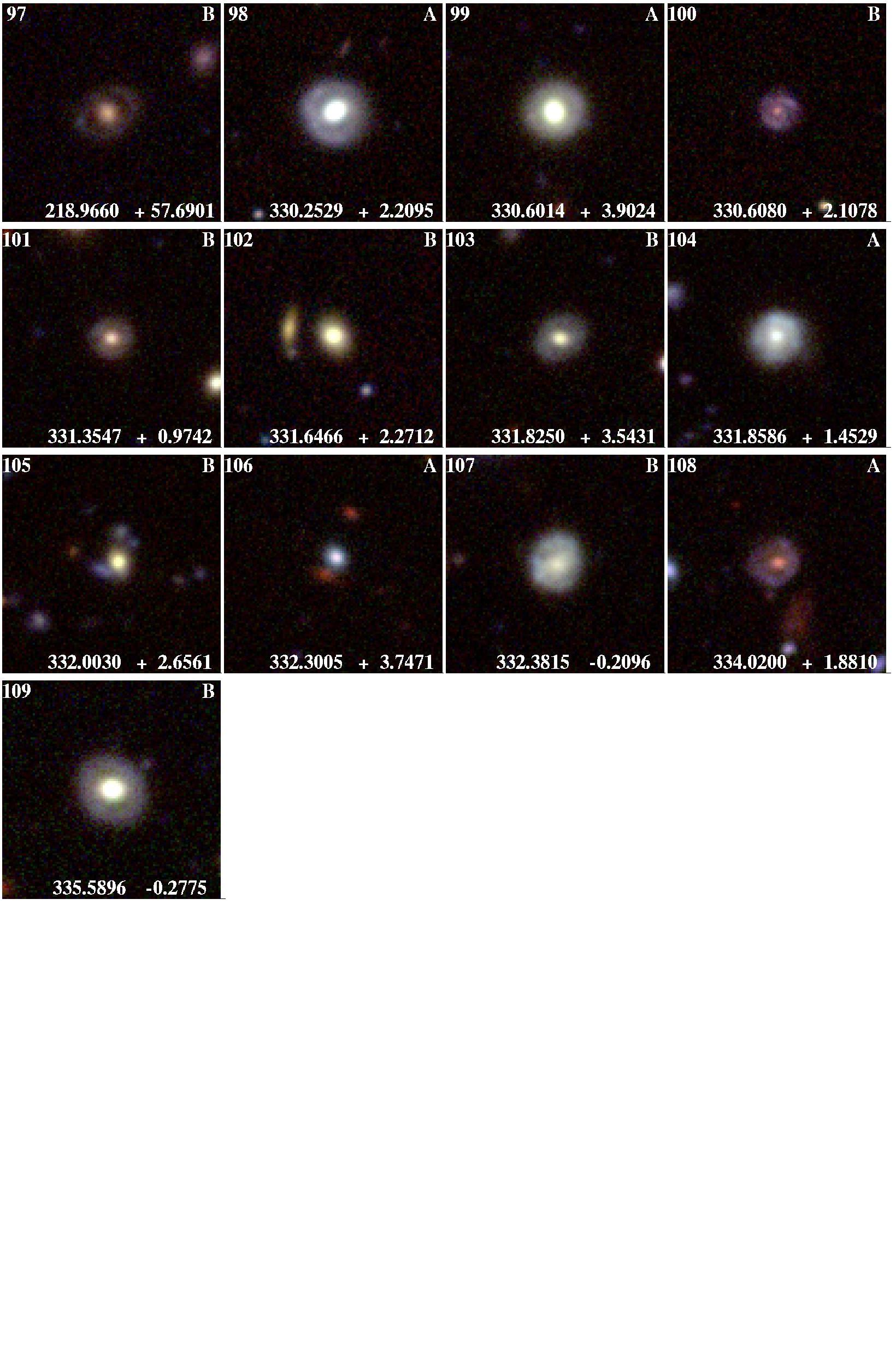}
 \caption{Our new lenses found with the PCA-finder, with grades $A$ and $B$ (see text). The stamps are 18.7\arcsec on-a-side.  \label{Fig5}}

\end{figure*}

In the following we provide colour stamps for our 109 lens candidates classified as grade-A or grade-B (Figs.~\ref{Fig1}-\ref{Fig5}). We also provide a list of the objects we classify as ring-like galaxies (Table~\ref{PolarRing}) or as a grade-C lens candidates (Table~\ref{QualityB}). 
\clearpage
\thispagestyle{empty}
\vspace{-1.5cm}
\begin{table*}
\scriptsize
 \begin{center}
\resizebox*{!}{\dimexpr\textheight-\lineskip\relax}{%
\begin{tabular}{rrl|lll|lll|llr}

\hline\hline
ID  & RA        & DEC        & ID  & RA        & DEC & ID  & RA        & DEC        & ID  & RA        & DEC     \\
\hline
    1 &30.7113&-4.1200 & 70&208.8451&  56.5793&139&212.0681&51.7086&207&216.0082&55.7079  \\
     2 &32.0140&-7.3349 & 71&208.8577&  55.2961&140&212.0702&54.6277&208&216.1492&53.3707  \\
     3 &32.7019&-11.091 & 72&208.8897&  56.7042&141&212.0781&53.1058&209&216.1771&56.1187  \\
     4 &32.9684&-10.640 & 73&208.9984&  57.6036&142&212.1310&56.5988&210&216.2110&53.6238  \\
     5 &33.3126&-4.5772 & 74&209.1030&  54.2619&143&212.1942&52.2143&211&216.3065&57.5956  \\
     6 &34.0313&-6.8846 & 75&209.1158&  53.3728&144&212.2530&53.6661&212&216.3152&54.6290  \\
     7 &34.6449&-6.2790 & 76&209.1330&  57.7878&145&212.3300&52.7731&213&216.4620&56.7888  \\
     8 &35.3084&-9.6229 & 77&209.1440&  53.1748&146&212.3553&55.9526&214&216.6035&54.0957  \\
     9 &36.2642&-6.6648 & 78&209.1640&  57.7423&147&212.4420&52.1788&215&216.6826&52.8987  \\
    10 &37.7604&-5.9290 & 79&209.1907&  56.3655&148&212.4687&51.9757&216&216.9283&57.2185  \\
    11&132.1819&-2.5873 & 80&209.2600&  57.224 &149&212.4777&56.5225&217&217.1591&55.7337  \\
    12&132.2712&-2.7720 & 81&209.2658&  53.3428&150&212.6073&56.4597&218&217.2200&55.6851  \\
    13&132.5795&-3.1395 & 82&209.3485&  54.5476&151&212.6626&57.4131&219&217.2537&55.4910\\
    14&132.5827&-2.1625 & 83&209.3499&  54.3683&152&212.6708&52.8456&220&217.3800&55.4627  \\
    15&132.5857&-4.6938 & 84&209.4350&  56.9008&153&212.8555&55.0499&221&217.4268&53.9160   \\
    16&132.6970&-3.9655 & 85&209.4560&  56.9913&154&212.9167&54.0741&222&217.4859&51.7736  \\
    17&132.7400&-1.0218 & 86&209.5048&  57.6376&155&212.9404&57.5752&223&217.5249&52.7104  \\
    18&132.7620&-4.4793 & 87&209.6050&  56.5502&156&212.9745&55.3304&224&217.5592&56.9327  \\
    19&132.9841&-1.9516 & 88&209.6506&  56.1549&157&213.0125&51.3368&225&217.6319&57.7677  \\
    20&133.0210&-4.6829 & 89&209.7750&  53.7688&158&213.1040&53.7781&226&217.7050&55.9099  \\
    21&133.1227&-1.7640 & 90&209.8462&  56.0677&159&213.1345&52.4077&227&217.8106&57.2403  \\
    22&133.1827&-1.0132 & 91&209.8569&  53.1140&160&213.1900&53.6537&228&217.9196&54.5902  \\
    23&133.2438&-3.7043 & 92&210.0242&  53.2755&161&213.2580&53.1980&229&218.0436&54.5195  \\
    24&133.5939&-4.9943 & 93&210.0400&  54.7716&162&213.2939&57.2640&230&218.0770&54.9706  \\
    25&133.6931&-2.7340 & 94&210.0919&  52.5197&163&213.3095&52.4831&231&218.3619&55.3975  \\
    26&133.7100&-5.0446 & 95&210.1264&  54.9506&164&213.5020&56.2184&232&218.5645&52.6184  \\
    27&133.7260&-3.9403 & 96&210.2546&  52.4485&165&213.5440&57.3132&233&218.6160&57.6086  \\
    28&133.7880&-1.4389 & 97&210.2719&  57.5983&166&213.5494&52.3798&234&218.6570&57.5602  \\
    29&133.9045&-4.3638 & 98&210.3997&  57.7682&167&213.7320&52.3263&235&218.8210&55.4744  \\
    30&133.9124&-2.3331 & 99&210.4121&  53.2592&168&213.7877&53.2259&236&218.8813&54.4891  \\
    31&133.9648&-4.7568 &100&210.4300&  51.2151&169&213.8380&57.1487&237&218.9065&51.3609  \\
    32&134.1370&-3.0186 &101&210.4339&  55.3332&170&213.9270&56.0135&238&219.3108&54.4433  \\
    33&134.1507&-1.3852 &102&210.4797&  51.2445&171&213.9639&51.3241&239&219.4381&56.3716  \\
    34&134.2170&-3.9283 &103&210.4847&  57.6922&172&213.9691&51.9893&240&219.5293&53.7316  \\
    35&134.2991&-4.3455 &104&210.4876&  53.2545&173&213.9950&51.5452&241&219.6096&52.1294  \\
    36&134.4163&-2.7038 &105&210.4968&  55.0219 &174&214.0239&57.0921&242&219.6133&52.4955  \\
    37&134.4683&-3.6448 &106&210.5020&  53.1556&175&214.1700&54.0163&243&220.2990&57.7254  \\
    38&134.5020&-3.2156 &107&210.5154&  52.3060&176&214.2040&55.4237&244&330.1120&1.5701  \\
    39&134.5101&-3.6909 &108&210.6305&  51.6363&177&214.4082&57.2170&245&330.2540&3.6684  \\
    40&134.5972&-2.1057 &109&210.7200&  52.8003&178&214.4454&55.4881&246&330.2720&1.8044  \\
    41&134.7143&-3.3857 &110&210.7309&  52.3358&179&214.5368&53.1975&247&330.4520&3.6380  \\
    42&134.8220&-4.1856 &111&210.7704&  53.7681&180&214.6275&56.8027&248&330.5939&1.2562  \\
    43&134.9660&-4.5596 &112&210.8229&  51.8781&181&214.6400&55.6239&249&330.8630&4.4709  \\
    44&135.0381&-3.3306 &113&210.8282&  56.4585&182&214.6506&54.2784&250&330.8980&2.2835  \\
    45&135.1403&-3.8596 &114&210.9047&  53.0148&183&214.6966&56.8438&251&331.3460&1.3502  \\
    46&135.3235&-5.5642 &115&210.9170&  53.8914&184&214.7002&53.3850&252&331.6890&1.0121  \\
    47&135.3369&-1.1031 &116&211.0441&  52.8848&185&214.7420&56.4807&253&331.8380&2.8518  \\
    48&135.3720&-1.5786 &117&211.0709&  57.2175&186&214.8210&56.2483&254&331.9330&2.3048  \\
    49&135.5498&-1.4060 &118&211.1030&  52.1859&187&215.0682&56.3292&255&332.0440&3.5210  \\
    50&135.5651&-2.0581 &119&211.1110&  57.0090&188&215.0730&56.531 &256&332.1489&1.9636  \\
    51&135.5864&-2.2146 &120&211.1478&  52.4499&189&215.0828&54.9762&257&332.2340&3.9315  \\
    52&135.8341&-1.1346 &121&211.2522&  52.0962&190&215.2030&56.333 &258&332.3800&0.5319  \\
    53&135.9001&-2.1360 &122&211.2820&  56.6721&191&215.2367&52.8727&259&332.6879&1.7517  \\
    54&135.9940&-3.2371 &123&211.2950&  53.9345&192&215.2490&51.4651&260&333.1200&0.8459  \\
    55&136.0742&-4.1396 &124&211.4487&  57.3926&193&215.3003&57.7346&261&333.1270&2.4188  \\
    56&136.1734&-2.0613 &125&211.5459&  56.7882&194&215.3255&57.5031&262&333.1709&0.3183  \\
    57&136.2500&-4.7148 &126&211.5520&  56.0455&195&215.3361&55.3473&263&333.2590&-0.8181  \\
    58&136.2506&-2.2494 &127&211.6047&  51.6347&196&215.3542&57.4317&264&333.4410&0.4691  \\
    59&136.3568&-2.6909 &128&211.6335&  52.0406&197&215.4756&56.7486&265&333.9400&1.5399  \\
    60&136.5063&-5.1453 &129&211.7230&  54.7437&198&215.5070&56.9580&266&334.1099&1.0777  \\
    61&136.5580&-1.1363 &130&211.7706&  54.6133&199&215.6347&54.5268&267&334.3880&1.2319  \\
    62&136.5629&-4.7081 &131&211.7880&  53.3319&200&215.6390&51.7208&268&334.4719&1.3240  \\
    63&136.5651&-2.1079 &132&211.8378&  52.5960&201&215.7157&55.2452&269&334.4760&2.7216  \\
    64&136.6360&-4.9990 &133&211.8520&  54.4707&202&215.8014&57.2830&270&334.6690&-0.4942  \\
    65&136.6770&-1.4028 &134&211.8555&  53.8529&203&215.9419&51.6791&271&335.2009&1.1618  \\
    66&136.7488&-1.6531 &135&211.8660&  54.7334&204&215.9590&51.6543&272&335.2869&0.1860  \\
    67&136.7710&-1.2799 &136&211.8971&  54.1434&205&215.9689&57.2974&273&335.3330&1.0047  \\
    68&208.6711&57.7116 &137&211.9844&  54.6894&206&216.0003&54.6786&274&335.5480&0.5849  \\
    69&208.7878&56.4358 &138&212.0647&  52.3376&&&\\

    \end{tabular}}
 \end{center}
 \caption{List of ring-like galaxies.\label{PolarRing}}

\end{table*}

\clearpage
\thispagestyle{empty}
\begin{table*}
\small
 \begin{center}
\begin{tabular}{ccr|ccr|ccr|ccr}
\hline\hline
ID  & RA        & DEC        & ID  & RA        & DEC  &  ID  & RA        & DEC        & ID  & RA        & DEC   \\
\hline
      1&  30.7690&  -4.3707&     47&  37.9131&  -8.4071&     93& 210.3239&  57.0069&    139& 213.6360&  53.4336\\
      2&  31.1471&  -6.8370&     48& 132.6063&  -1.6624&     94& 210.3239&  57.0069&    140& 213.8290&  51.5396\\
      3&  31.1845&  -9.2302&     49& 132.8670&  -1.7824&     95& 210.3830&  52.9747&    141& 213.8425&  54.5827\\
      4&  31.1953&  -7.4963&     50& 133.0932&  -5.5540&     96& 210.4185&  51.9295&    142& 213.8930&  55.9188\\
      5&  31.3286&  -9.4541&     51& 133.1093&  -2.1114&     97& 210.4440&  56.0728&    143& 214.0553&  56.3335\\
      6&  31.4334&  -5.5922&     52& 133.6085&  -3.3218&     98& 210.4970&  55.0210&    144& 214.1416&  54.2238\\
      7&  31.4347&  -8.9391&     53& 133.8673&  -4.4843&     99& 210.5960&  56.7669&    145& 214.1440&  52.1982\\
      8&  31.7222&  -6.9676&     54& 134.1219&  -2.8850&    100& 210.8169&  56.3686&    146& 214.2251&  53.2605\\
      9&  32.0885& -10.1001&     55& 134.3641&  -3.8366&    101& 210.8650&  54.0454&    147& 214.4718&  56.4726\\
     10&  32.1265&  -8.6989&     56& 134.4226&  -5.5544&    102& 210.9850&  53.6275&    148& 214.6783&  52.0068\\
     11&  32.1489& -10.6963&     57& 134.9147&  -1.7250&    103& 210.9875&  52.7897&    149& 214.8440&  52.0608\\
     12&  32.3864&  -8.6895&     58& 135.0480&  -4.2772&    104& 211.0765&  56.2987&    150& 214.8870&  55.7473\\
     13&  32.5458&  -6.9613&     59& 135.3490&  -2.7373&    105& 211.0960&  52.7181&    151& 214.9260&  56.2809\\
     14&  32.6089&  -4.2655&     60& 135.3985&  -4.6624&    106& 211.1021&  56.0841&    152& 215.2367&  52.8727\\
     15&  32.6312&  -5.3873&     61& 135.4108&  -4.9603&    107& 211.2520&  52.0962&    153& 215.5535&  52.0773\\
     16&  32.7019& -11.0916&     62& 135.5910&  -2.0328&    108& 211.2966&  53.9393&    154& 215.5858&  52.3223\\
     17&  32.7654& -10.1586&     63& 135.7976&  -3.1933&    109& 211.2975&  52.6984&    155& 215.5870&  52.5406\\
     18&  32.7713&  -4.3339&     64& 135.8190&  -1.4759&    110& 211.4610&  56.5435&    156& 215.9215&  53.1101\\
     19&  32.7958&  -9.1606&     65& 135.8390&  -4.7139&    111& 211.5791&  51.5461&    157& 216.0830&  56.5382\\
     20&  32.8151&  -4.6442&     66& 135.8770&  -3.3166&    112& 211.6760&  56.8844&    158& 216.1280&  53.7346\\
     21&  32.8263&  -5.9572&     67& 135.8916&  -4.7350&    113& 211.7245&  54.9516&    159& 217.1294&  53.2367\\
     22&  33.1106&  -9.1819&     68& 136.1790&  -2.0222&    114& 211.7933&  57.7058&    160& 217.2020&  57.1215\\
     23&  33.1371&  -8.2071&     69& 136.3051&  -1.7993&    115& 211.8378&  52.5960&    161& 217.3254&  51.2955\\
     24&  33.4489&  -5.0069&     70& 136.4287&  -4.3918&    116& 211.8980&  54.3523&    162& 217.3837&  57.0770\\
     25&  33.7189& -10.2549&     71& 136.6511&  -4.1206&    117& 211.9132&  56.5606&    163& 217.4874&  53.4669\\
     26&  33.8125&  -7.6329&     72& 136.6759&  -3.4018&    118& 212.0111&  54.5693&    164& 217.5479&  53.9042\\
     27&  33.8846&  -7.3768&     73& 136.7700&  -3.6983&    119& 212.3363&  53.7088&    165& 218.5887&  53.5343\\
     28&  34.0291& -10.4792&     74& 208.5923&  56.9917&    120& 212.5192&  52.8386&    166& 218.9180&  51.5978\\
     29&  34.0313&  -6.8846&     75& 209.0410&  55.1548&    121& 212.5344&  53.5059&    167& 218.9306&  51.6789\\
     30&  34.0420&  -4.9278&     76& 209.0520&  55.3643&    122& 212.5578&  57.1932&    168& 218.9306&  51.6789\\
     31&  34.2354&  -7.3950&     77& 209.0883&  54.6384&    123& 212.6180&  54.6689&    169& 219.2150&  53.1183\\
     32&  34.4127&  -5.6183&     78& 209.1308&  56.8001&    124& 212.7260&  56.5193&    170& 219.2270&  54.7119\\
     33&  34.7033&  -6.3146&     79& 209.3442&  56.0257&    125& 212.7260&  54.2777&    171& 219.2620&  53.0397\\
     34&  34.7611&  -5.6861&     80& 209.4020&  56.8134&    126& 212.7260&  56.5193&    172& 219.4122&  56.8565\\
     35&  34.7705&  -8.1057&     81& 209.6070&  52.7409&    127& 212.7370&  53.3133&    173& 219.5995&  53.3351\\
     36&  35.1727& -10.8737&     82& 209.6217&  52.6670&    128& 212.8499&  56.3970&    174& 219.6419&  54.4867\\
     37&  35.4835& -10.6293&     83& 209.6358&  56.3116&    129& 212.8877&  56.1517&    175& 219.6730&  56.9740\\
     38&  35.4901&  -6.8625&     84& 209.8516&  54.3894&    130& 212.9065&  53.8578&    176& 331.1035&   1.3934\\
     39&  35.5342&  -8.2175&     85& 209.9950&  56.2114&    131& 212.9810&  52.5545&    177& 331.8447&   4.2985\\
     40&  36.0965&  -3.9120&     86& 210.0480&  53.4897&    132& 213.1729&  54.4224&    178& 331.9094&   1.6015\\
     41&  37.1595&  -8.0260&     87& 210.1192&  56.0877&    133& 213.1870&  55.2213&    179& 332.0385&   1.5294\\
     42&  37.3517& -11.1634&     88& 210.1260&  54.9506&    134& 213.3750&  53.4311&    180& 332.0615&   2.6110\\
     43&  37.3725&  -9.9510&     89& 210.1920&  55.5750&    135& 213.4316&  53.0792&    181& 332.1574&   3.3451\\
     44&  37.7770& -10.0155&     90& 210.2130&  52.9077&    136& 213.5034&  55.1556&    182& 332.7543&   0.0248\\
     45&  37.8688&  -9.2495&     91& 210.2437&  56.8107&    137& 213.5379&  52.4758&    183& 333.8364&   0.9369\\
     46&  37.9131&  -8.4071&     92& 210.3010&  57.0831&    138& 213.5680&  54.4716&
    \end{tabular}
\end{center}
  \caption{List of our grade-$C$ candidates in CFHTLS. \label{QualityB}}
\end{table*}
\clearpage
\thispagestyle{empty}

\clearpage
\thispagestyle{empty}
\clearpage
\thispagestyle{empty}

\clearpage
\thispagestyle{empty}

\end{appendix}

\end{document}